\documentclass[11pt]{revtex4}
\usepackage{epsf}
\usepackage{graphicx}
\usepackage{dcolumn}
\usepackage{bm}
\def\prb{Phys. Rev. B}
\def\prl{Phys. Rev. Lett.}

\def\be{\begin{equation}}
\def\ee{\end{equation}}
\def\ba{\begin{eqnarray}}
\def\ea{\end{eqnarray}}

\begin{document}

\draft

\title{Transport in Two Dimensional Electronic Micro-emulsions}

\author{Boris Spivak}
\affiliation{Department of Physics, University of Washington,
Seattle, WA 98195}

\author{Steven A. Kivelson}
\affiliation{Department of Physics, Stanford University, Stanford, CA 94305-4045}

\begin{abstract}
In two dimensional electron systems with  Coulomb or dipolar interactions,
a direct transition, whether first or second order, from a liquid to a  crystalline state is forbidden.
As a result, between these phases there must be other (microemulsion) phases
which can be viewed as a meso-scale mixture of the liquid and crystalline phases.
We investigate the transport properties of these new electronic phases and
  present
arguments that they are responsible for the various
transport anomalies that have been seen
 in experiments on the strongly correlated 2DEG in high mobility semiconductor devices with low electron densities.
 \end{abstract}


\maketitle

As low density  two dimensional (2D) electronic systems with
 increasingly high mobility have become available, there has accumulated
  experimental evidence of a set of low temperature
   phenomena   which cannot be understood on the basis of
   traditional (Fermi liquid based) metal physics
  \cite{AbrKravSar,pudalov,krav,sar,kravHpar,pudHpar,okamoto,pudanis,VitSat,Gershenzon,Shahar,GaoLanl,
Gao2005,Gao2002,GaoHpar05,Pillarisetty,gao01}.
     (More precisely, by low density we mean a large  ratio between the potential and
     the kinetic energy in the liquid measured by the parameter
     $r_s  \sim 1/\sqrt{n(a_B)^2}$,
 where $n$ is the areal density of electrons and $a_B$ is the effective Bohr radius.)
A generic feature of electronic systems is a low temperature
evolution as a function of $n$ from a conducting ``Fermi
liquid'' (FL) state at large $n$ to an insulating ``Wigner crystalline'' state (WC) at low.   Much is understood about these limiting cases, so it is possible to make rather firm statements about the failure of these states to account for the experimental features.

Recently, we
{\cite{spivakconf,SpivakPS,SpivakKivelsonPS,Reza,SpivakKivelsonDrag}
have shown that  in ideal 2D systems there must exist new electronic
micro-emulsion
 phases at intermediate densities -- phases which consist of various sorts of
 micro-phase separated mixtures of regions of WC and regions of FL.  This follows from
  the fact that, in the presence of
  either
  Coulomb or dipolar
  interactions, neither a direct first or second order transition from a uniform liquid to an
  electronic crystal is allowed.
  We have argued that these micro-emulsion phases may be responsible for the anomalous
behaviors observed in experiment.

Electronic micro-emulsion
phases are distinct
thermodynamic phases of matter;
transitions between them are associated with  nonanalytic
dependencies of the free energy  on
$T$ and $n$.
Since different micro-emulsion phases exhibit different patterns
of phase separation on the mesoscale, they also have new
forms of hydrodynamics. For example,
when the
minority phase consists of dilute bubbles of
FL
embedded in a WC host,
the result is
a relative of a supersolid phase \cite{andreevSupersolid}:
the hydrodynamics of such a state combines usual hydrodynamics with elasticity theory.
The hydrodynamic properties of other microemulsion phases are even more
complicated both on the classical and  quantum levels.
On the classical level, some of them are similar (but not
identical) to
classic liquid crystals.

 It is the purpose of the present paper to report a step
toward a theory of the low energy, long
wave-length dynamical properties of these micro-emulsion phases,
and to begin the exploration of how they are modified in the
presence of weak quenched disorder. 
We  have also
   undertaken to compare those
   aspects of the theory which we do understand with
    experiments on the 2D electron gas (2DEG) in
     Si-MOSFETs \cite{AbrKravSar,pudalov,krav,sar,kravHpar,pudHpar,okamoto,pudanis,VitSat}, and the
     newer experiments on the 2D hole gas (2DHG) in GaAs heterostructures \cite{Shahar,GaoLanl,
Gao2005,Gao2002,GaoHpar05,Pillarisetty,gao01}.

 Since this paper is long,
efforts have been made to make the various sections
  self-contained.
 Each section begins with a summary
    of the important results.
In Section \ref{experiment}, we discuss some of the features of the observed behaviors
 of the 2DEG and 2DHG at large $r_s$ which we believe are incompatible with a Fermi
  liquid description.  In  Section \ref{thermo} and Appendix \ref{thermoappend}, we review the thermodynamic analysis
   which implies the existence of electronic micro-emulsion phases. Much of this discussion is review, although
     we have expanded considerably upon previous results.  The central result
  in this discussion is Eq. \ref{Fmicro1} (or its more general
   form in Eq. \ref{Fmicro}), which expresses the free energy density of a
    general micro-emulsion phase in terms of a non-local surface tension.
    Section \ref{beyond} is a slight digression, in which the relation is derived between hydrodynamics of a fluid in the absence of disorder and the resistivity in the presence of
    weak disorder.
In Section \ref{bubble} we discuss the finite $T$ hydrodynamics of what is in many ways
        the simplest microemulsion phase, a WC bubble fluid consisting of a dilute
         concentration of WC bubbles in a majority FL sea.
   In Section \ref{quantumbubble} and Appendix \ref{delocalization} we analyze the  low $T$ limit where the quantum nature of the bubbles begins to play an important role.  In Section \ref{complex} we briefly discuss the hydrodynamics of some of the other microemulsion phases.
   In Section \ref{drag} we consider the
    case in which there is a bilayer of two 2DEG's which permits measurements of  the ``drag resistance''.   In Section \ref{comparison}, we compare the results obtained in the present paper
   with corresponding existing experimental observations on the 2DEG and 2DHG.
We also discuss what we think are the major shortcomings of some of the alternative,
    Fermi liquid based explanations of the same data that have been discussed in the
     literature.
       Finally, in Section \ref{conclusion}, we conclude with a discussion of a number of important
       remaining open issues.

 Throughout this article, we  work in units in which Boltzman's constant, $k_B=1$, and the Zeeman splitting, $g\mu_B\hbar=1$, where $g$ and $\mu_B$ are, respectively, the gyromagnetic ratio and the effective Bohr magneton of an electron.

\section{Experimental signatures of Non-Fermi liquid behavior in strongly correlated 2DEGs}
\label{experiment}

In this section, we summarize some of the experimental results on large $r_s$ 2DEG's
 and 2DHG's that we think are incompatible with
 Fermi liquid based theory. We  mainly focus on
  the behavior of systems with large $r_s$
  but also small values of the resistivity $\rho$ (compared to the quantum of resistivity
   $h/e^2$), since in this limit, the predictions of Fermi liquid theory are crisp,
    and so discrepancies with the theory can be documented.
(A  more specific discussion of some  proposed FL based theories can be found in Section \ref{sarma}.)

To orient the reader, the devices we have in mind have $r_{s}\sim 5-20$ in Si MOSFETs and $r_{s}\sim 10-40$ in p-GaAs.   In comparison, the best current theoretical estimates \cite{cip} of the critical value of $r_s$ at which the energies of a uniform FL and WC are equal is $r_s^\star \approx 38$.

\subsection{Experiments on single layers.}

 {\bf a}. {\it The metal-insulator transition:}
Both the 2DEG and the 2DHG  exhibit an ``apparent zero-temperature metal-insulator
 transition'' as a function of the electron density $n$, or the gate voltage in
MOSFET's. (See for example, Fig. 1 in Ref. \cite{AbrKravSar} ).
 Applying a magnetic field,
$H_{\|}$, parallel to the 2D plane
 drives the system toward the insulating
phase. Thus the critical
concentration $n_{c}(H_{\|})$ at which the transition occurs increases with
$H_{\|}$.
The existence of such a transition is, perhaps, the most fundamental signature
 of the breakdown of
 the Fermi liquid based theory of localization,
 since in this framework there is no metallic phase in the 2DEG at $T \to 0$.
 While it is difficult to establish unambiguously that the apparent transition is not actually a rapid
  crossover,  the resistivity certainly changes by several orders of magnitude
as  $n$ varies over a modest range near a critical concentration. The range of resistances and temperatures covered
   by the experiments is, in some cases, impressively large.

Where  the resistance $\rho\ll h/e^{2}$,
were FL theory applicable, perturbation theory in powers of
    $1/k_F\ell$ would be well controlled.  Thus it is  relatively straightforward to identify
     behaviors that are {\it not} consistent with FL theory in the ``metallic'' regime.  (Here, and henceforth, we use the term ``FL theory'' in a somewhat expanded sense to refer to results based on FL theory plus low order perturbative corrections of various sorts in powers of $1/k_F\ell$.)

 {\bf b}. {\it Temperature dependence of the resistitivity $\rho(T)$ in strongly correlated ($r_{s}\gg 1$)
  metallic ( $\rho\ll  h/e^{2}$) samples:}
 At low temperatures, $\rho(T)$ is observed to increase with increasing temperature  by as much as a factor of six  in
Si-MOSFET's (See  Fig. 1 in Ref. \cite{AbrKravSar} ) and
by a  factor of three or more in GaAs heterostructures (See Fig. 1 of Ref.\cite{GaoLanl} and Fig. 3 of Ref. \cite{Gao2005}).
At $T=T_{max}\sim E_{F}$ the resistance has a maximum at
 which $h/e^{2}\gg \rho_{max}\gg \rho(T=0)$; then at $T>T_{max}$ it decreases
with $T$. The resistivity maximum is not terribly pronounced in the data
on Si MOSFET's, but some of the data  taken in  p-GaAs exhibit this maximum   very clearly, as can be seen in Fig. 1 of Ref.\cite{GaoLanl} and Fig. 3 of Ref. \cite{Gao2005}.
In the latter case $\rho \ll h/e^2$ in the entire range of the temperatures of interest
so interference corrections to the conductance are unimportant.

 Such a strong temperature dependence of the resistance is
unexpected in the context of Boltzman transport in a FL at low $T$
  where electron-phonon scattering is negligible.
(In many cases, a crossover temperature, $T_{ph}$ below which electron-phonon scattering is unimportant can be readily identified; for example, in Fig. 3 of Ref. \cite{Gao2005},
$T_{ph} > E_{F}$ is roughly the point at which $\rho(T)$ has a minimum.)
 Since the Fermi momentum is much smaller than the reciprocal
lattice vector, electron-electron scattering conserves the
total quasi-momentum and, therefore does not contribute directly to the
resistance.
 The resistance is therefore
determined by the rate of electron-impurity scattering, which typically is a weakly
 {\it decreasing} function of the temperature in a FL.

{\bf c}. {\it  Parallel field magneto-resistance of strongly correlated ($r_{s}\gg 1$)
  metallic ( $\rho\ll h/e^{2}$) samples:}
  At
$T<E_{F}$ and at sufficiently large $r_{s}$, both in Si
MOSFET's (See Fig. 3 in Ref. \cite{kravHpar} and Figs. 1 in Ref. \cite{pudHpar,pudanis}) and GaAs hetero-junctions
(See Fig. 1 in Ref. \cite{GaoHpar05}), the 2DEG  exhibits a strongly
{\it positive} magneto-resistance as a function of $H_{\|}$, which
saturates at $H_{\|}>E_{F}$.

 In sufficiently thin samples, $H_{\|}$
has little effect on the orbital motion of the electrons,
 so
 it can be viewed as coupling only to the electron spin,
  and therefore varies the degree of spin polarization of the electron liquid.
    At $T=0$, for fields in excess of a critical field, $H^{\star} \sim E_{F}$, the electron
     gas is fully polarized, and consequently $E_F$ is increased to twice its zero
      field value. The low temperature
magneto-resistance in this case comes from a combination of the
$H_{\|}$ dependencies of the Fermi velocity, the density of states
and the electron-impurity cross-section.
In a FL, the doubling of $E_F$ typically should lead
  to a
  {\it decrease} in the resistivity
 because scattering cross-sections are generally decreasing functions of energy.

{\bf d}. {\it $H_{\|}$ dependence of the slope $K\equiv d \rho(H_{\|})/ dT$:}
 At large magnetic field ( $
H_{\|}>
H^\star $), the slope $d\rho(T)/dT$ is significantly smaller than when 
$H_{\|}=0$  -- by as much as a factor 100! This effect has been observed in Si MOSFET's
(See Fig. 3 in Ref.
 \cite{VitSat}) and in p-GaAs (See Fig. 2 in Ref. \cite{GaoHpar05}).
 In the FL framework
 there
is no reasons for such a dramatic effect.
Any scattering process
 responsible for the  $T$ dependence of the resistivity at $H_{\|}=0$ would typically
  be expected to give rise to $T$ dependent scattering even when the electron gas is
   fully polarized.

 {\bf e.} {\it ``Metallic'' $T$ dependence in samples with large (``insulating'') resistivity, $\rho > h/e^2$:}
 The conventional theory  of localization in the strong disorder limit $\rho \gg h/e^2$ predicts that the
  electronic transport should be due to hopping conductivity.
 Therefore the resistivity of such conductors should increase with {\it decreasing} temperature
 and diverge as $T\rightarrow 0$. On the other hand, in some  p-GaAs samples (See for example Fig.1 in
 Ref. \cite{gao01}) with large $r_s$, the low temperature resistivity exhibits a ``metallic'' temperature
  dependence, {\it i.e.} $\rho$ increases with {\it increasing } temperature, even when $\rho > h/e^2$.

{\bf f.}  {\it Oscillations of  $\rho(H_{\bot})$ at $T>E_{F}$ }.  Oscillations of the resistivity, $\rho(H_{\bot})$ have been observed as a function of the strength of a perpendicular magnetic field, $H_{\bot}$
in both the Si MOSFET's
and in
p-GaAs heterojunctions.
At relatively small $r_{s}$, these
oscillations were interpreted as Shubnikov-de Haas oscillations.
However, for samples with large $r_s$ reported in
Ref. \cite{GaoLanl}, these oscillations persist
up to temperatures which are significantly larger than the bare Fermi
energy.  It is important to stress that FL theory not only predicts the existence of magnetic oscillations
 with period inversely proportional to the area enclosed by the Fermi surface, it also predicts that
  these oscillations should be entirely quenched
when $2\pi T$ exceeds the cyclotron energy, $\omega_c=eB/mc$, which in turn must be less than or equal to $E_F$.  Thus, the persistence of the oscillations
to such high temperatures  is in disagreement with Fermi liquid
theory !

\subsection{Experiments on double-layers}

Additional information concerning correlation effects can be obtained from measurements of the ``drag''
 resistance in  a system of two 2DEG layers which are electrically unconnected.
Current 
$I$ is passed through the lower (active) layer and  the voltage $V_{D}$ is measured on
the upper (passive) layer.
The drag resistance is defined to be the ratio $\rho_{D}=V_{D}/I$.

For relative small $r_{s}$, experiments \cite{Eisenstein} on the drag resistance,
 $\rho_{D}(T)$, in
double layer 2DEGs are in qualitative agreement with Fermi liquid
theory \cite{Price,MacDonnald}. Specifically, the drag resistance is small in
 proportion to $(T/E_F)^2$ and in proportion to $(k_Fd)^{-\alpha_d}$ where $k_F$ is
  the Fermi momentum, $d$ is the spacing between the two layers, and typically
   $\alpha_d = 2$ or 4, depending on the ratio of $d/\ell$. However,   experiments
    \cite{Pillarisetty}  on p-GaAs double layers with $r_{s} \sim 20-30$,
 yield results which differ significantly from the predictions of Fermi liquid theory.
  These experiments are performed on samples with small resistances $\rho\sim 0.05 - 0.1\  h/e^2$,
in which quantum interference corrections to the Drude conductivity are insignificant:

{\bf a.} {\it The magnitude of the drag resistance:}  The drag resistance in these samples
is 1-2 order of
magnitude larger than expected on the basis of Fermi liquid
theory.

{\bf b.}  {\it $T$ dependence of $\rho_{D}(T)$:}  Whereas in a Fermi liquid, $\rho_{D}(T)\sim T^2$,
 in large $r_s$ devices $\rho_{D} \sim (T)^{\alpha_T}$ where
the temperature exponent  exhibits non-Fermi liquid values
$2 < \alpha_{T} < 3$.  (For example, $\alpha_{T}=2.7$ in Ref. \cite{Pillarisetty}.)

{\bf c.} {\it $H_{\|}$ dependence of $\rho_{D}(T, H_{\|})$:}
At low temperature, $\rho_{D}(T,H_{\|})$ {\it increases} as a function of $H_{\|}$ by a
factor of 10-20  and saturates when $H_{\|}>H^{*}$ (See Fig. 1 in Ref.
\cite{Pillarisetty} ). In contrast, in the framework of the Fermi liquid
theory, one would instead expect a significant {\it decrease} of $\rho_{D}(H_{\|})$
in the spin polarized state ( $H_{\|}>H^{\star}$) for the familiar reason that $E_F$ is
increased by a factor of 2 at $H_{\|}>H^{\star}$, and hence the
electron-electron scattering rate decreases.

A parallel magnetic field also appears to strongly suppress the temperature dependence of
$\rho_{D}(T)$ \cite{Pillarisetty}.
In the presence of non-zero $H_{\|}$, the value of
$\alpha_T(H_{\|})$ decreases
 with $H_{\|}$
 and saturates for $H>H^{\star}$ at a value which is significantly
  smaller
than the FL value $\alpha_T=2$.
 (For example, in Ref. \cite{Pillarisetty}
$\alpha_{T}(H>H^{*})\sim
 1.2$.)

{\bf d. }{\it The relation between $\rho(T, H_{\|})$ and $\rho_{D}(T, H_{\|})$:}
  The $T$ and especially the $H_{\|}$ dependencies
  of $\rho_{D}(T, H_{\|})$ and the resistivities of the individual
   layers $\rho(H_{\|}, T)$ look
qualitatively similar to one another, which suggests that both have a common
    origin. (See Figs.1 a,b in
\cite{Pillarisetty} ). In a Fermi liquid, $\rho(T, H_{\|})$ is primarily determined by the
 electron-impurity scattering, and $\rho_{D}(T, H_{\|})$ by the inter-layer
 electron-electron scattering, so there is no reason for their $T$ and
  $H_{\|}$ dependencies to be similar.

\subsection{Comparison with small $r_s$ devices}

 The anomalies discussed above have been observed  in samples with large $r_s$.
 In relatively smaller $r_s$ high mobility devices ( {\it i.e.} $r_s \sim 1$), behavior much more in line
  with the expectations of FL theory, modified by weak interference effects
 (See, for example, Ref. \cite{stern} for a review and Ref. \cite{Gershenzon} for experiments  in Si MOSFET's with high electron mobility and large distance to the gate. )

\section{Thermodynamics of Electronic Micro-emulsions}
\label{thermo}

 The two well understood phases of the 2DEG are a kinetic energy dominated
 FL phase ($r_s \ll 1$) and an insulating
 WC phase ($r_s\gg 1$), which is a state of spontaneously broken translational symmetry.
 \footnote{We will not explicitly consider the case in some solids, where lattice commensurability plays an important role, and the corresponding state is known as a Mott insulator;  this case shares some features in common with the Wigner crystal, but has its own peculiarities, which must be treated separately.}
  Were there a direct transition from a WC to an isotropic liquid, there are compelling reasons to think it would be first order.  According to Landau, (See for example \cite{chaikin}.) liquid-crystal transitions are expected to be
 first order because of the presence of cubic invariants in the free energy functional.   Brazovskii
 \cite{brazovski} showed that, even in cases (such as the liquid to smectic transition) where there are no cubic invariants, a freezing transition is expected to be fluctuation driven first order.  However, we have proven (See below.) that, in general, no first order transition is possible when  there are long, or moderately long-range interactions: $V(r) \sim Q/r^x$ with $D-1\le x \le D+1$ where $D$ is the number of spatial dimensions.  This result applies, therefore, to both the Coulomb ($x=1$) and dipolar (x=3) cases in D=2.

  The apparent contradiction - that the transition must be first order and cannot be - is resolved by the existence of intermediate phases;  there is no direct transition between a Wigner crystal and a uniform fluid phase.  Rather, electronic microemulsions (ME), 
  which can be visualized as consisting of intermingling regions of the two extremal phases, occur at intermediate densities.
In essence, these
phases are the vestige of the two-phase coexistence that would have accompanied the putative first order transition.
The fraction of the system that is locally crystalline or fluid, as well as the size, shape and ordering of these regions are all determined by minimizing the free energy.
 It is important to stress that,  since the energy associated with the patterns is
extensive, these microemulsions constitute new
phases in the thermodynamic sense. \cite{note1}
 They  provide an
important paradigm of new non-Fermi liquid phases.
On the mean field level, ME phases have many features in common with domain structures in 2D uniaxial ferromagnets \cite{2Dferromagnets,Marchenko} and lipid films on the surface of water \cite{lipids}.  They are also similar to  the various micro-phase separated states found in transition metal oxides\cite{tokura,emeryphysicac,low,aeplipnas,fradkin,tranquada,howald,lang,pan,rmp,dagotto}
(including the stripe phases in the cuprate high temperature superconductors), the Bechaard salts\cite{yu,lee,vuletic}, and the  ET based organic superconductors\cite{lefebvre}.

 The mean field phase diagram for microemulsion phases has been discussed in Refs. ~\cite{SpivakPS,SpivakKivelsonPS,Reza}. This section therefore mostly describes the results, with  the derivations (which generalize our earlier work)  relegated to Appendix \ref{thermoappend}.  The central  result in the present section is Eq. \ref{Fmicro1}, which  gives the free energy density in the presence of either Coulomb or dipolar interactions of a microemulsion phase with domains of arbitrary shape and size.   The mean-field phase diagram is derived by minimizing this equation.  The resulting  
$T=0$ 
mean field phase diagram for the 2DEG in the dipolar case ({\it e.g.} a disorder-free  MOSFET) is shown in Figs. \ref{fig:fig1} and \ref{fig:fig3}. The phase diagram for the 2DEG without a ground-plane is obtained from this by taking the limit $d\to\infty$, where $d$ is the distance to the ground-plane.

At finite $T$, the Pomeranchuk effect implies that the fraction of WC {\it increases} with increasing temperature and $H_{\|}$, as discussed in Subsection \ref{Pomeranchuk}.  The resulting  phase diagram is shown in Figs. \ref{fig:fig4} and \ref{fig:fig5}.

\begin{figure}
  \centerline{\epsfxsize=10cm \epsfbox{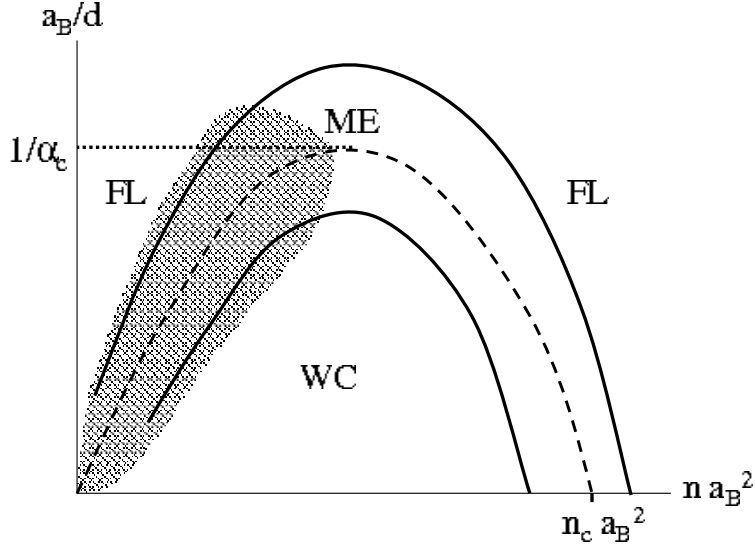}}
  \caption{ The $T=0$ phase diagram of the 2DEG in an ``ideal'' MOSFET, where $d$ is the distance to the ground-plane.
 $n_{c}(d)$ is the putative critical density at which 
   the free energies of the uniform Wigner crystal (WC) and Fermi liquid (FL) phases would be equal in the absence of microemusion (ME) phases.
The solid lines mark the boundaries of the regime of the
 intermediate ME (stripe or bubble) phases.
  At mean field level, these lines are Lifshitz transitions. The
  hatched area represents the regime in which the regions
  of the two coexisting phases have sizes of order the electron
  spacing, so quantum fluctuations are order 1.  
 $\alpha_c$ is the critical value of $d/a_B$ as discussed in Sec. \ref{ideal}.
  }  \
  \label{fig:fig1}
\end{figure}

\begin{figure}
  \centerline{\epsfxsize=10cm \epsfbox{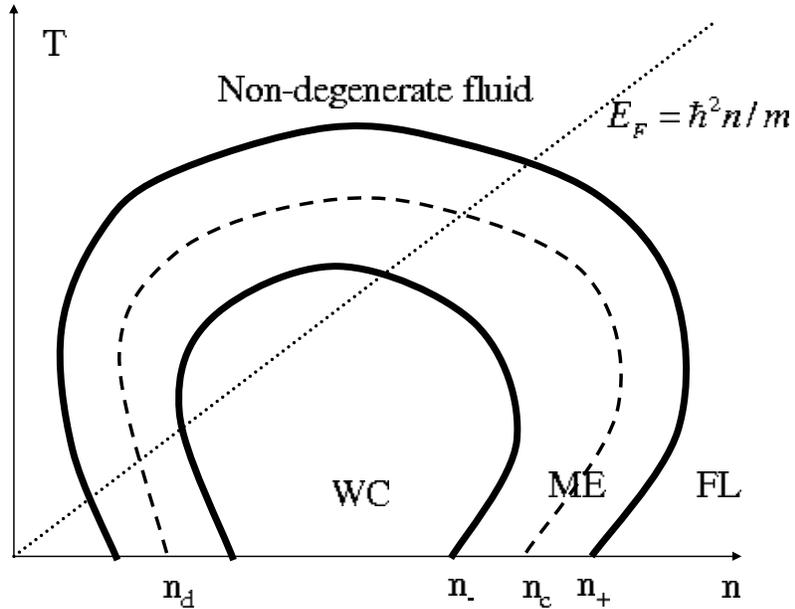}}
  \caption{ Mean field phase diagram of the 2DEG in an ideal MOSFET in the  $T \ - \ n$ plane.
  The dashed line indicates the putative phase transition between uniform FL and WC phases.
  The dotted line indicates the $n$-dependence of the bare Fermi energy.
 ME stands for microemulsion phases.  $n_d$ is the lower critical density (proportional to $d^{-2}$ where $d$ is the spacing to the ground plane) at which the putative reentrant WC to FL transition occurs.
  } \
  \label{fig:fig4}
\end{figure}

\begin{figure}
  \centerline{\epsfxsize=10cm \epsfbox{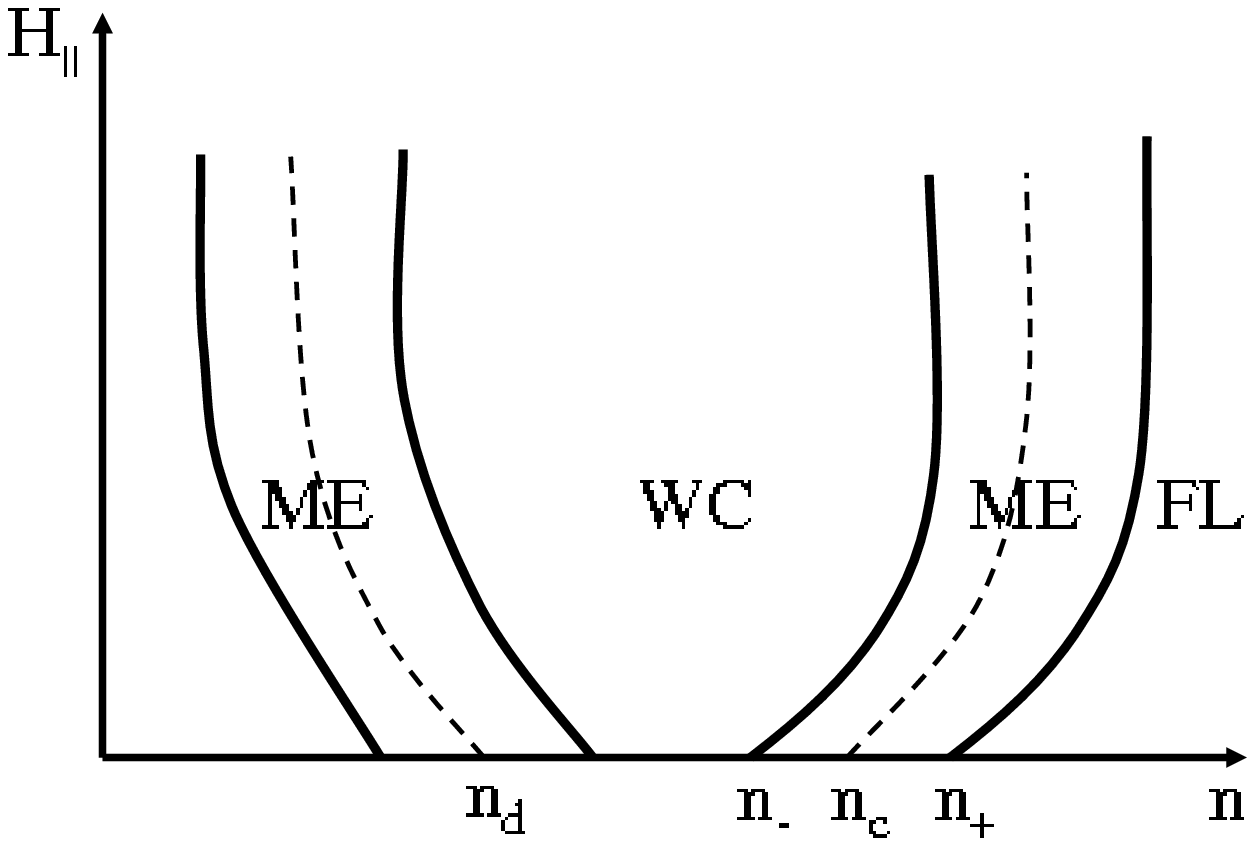}}
  \caption{ Mean field phase diagram of the 2DEG in an ideal MOSFET
  in the $H_{\|} \ - \ n$ plane.
  The dashed line indicates the putative phase transition between uniform the FL and WC phases.
  Symbols are is in Fig. \ref{fig:fig4}
  } \
  \label{fig:fig5}
\end{figure}

\subsection{Mean-Field theory of Electronic Microemulsions}

A putative first order transition from the WC to the FL would occur when the free energy densities of the two uniform phases, $F_{WC}(n)$ and $F_{FL}(n)$, cross at a critical density, $n_c$, as in Fig.  \ref{fig:fig2}.    
Recall that in the case of sufficiently short-range interactions, where such a transition can truly happen, it  proceeds somewhat differently if considered as a function of chemical potential, $\mu$, or  of the mean density, $\bar n$:  As  $\mu$ is varied through the critical value, the system jumps from a uniform WC with $\bar n=n_-$, to a uniform FL with  $\bar n=n_+$, where $n_\pm$ are determined from $F_{WC}(n)$ and $F_{FL}(n)$ by the Maxwell construction.  Alternatively, as a function of $\bar n$, the equilibrium state with $n_-<\bar n < n_+$ is macroscopically phase separated into two regions with density $n_-$ and $n_+$ and areal fractions $(n_+-\bar n)/(n_+-n_-)$ and $(\bar n-n_-)/(n_+-n_-)$, respectively.

\begin{figure}
  \centerline{\epsfxsize=10cm \epsfbox{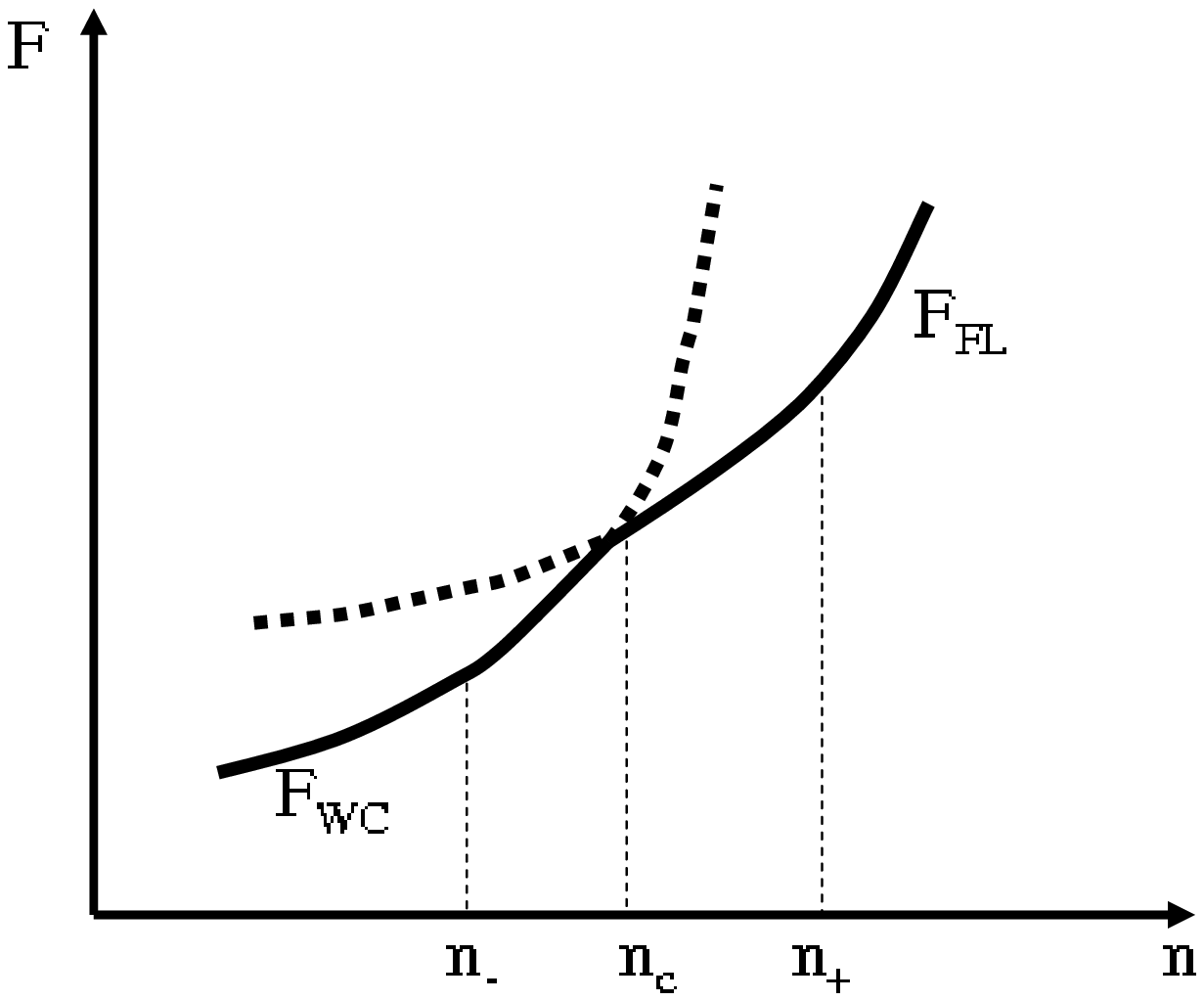}}
  \caption{Mean-field  $n$- dependences of the Free energy densities of the uniform FL and WC phases in the dipolar case.  As is well known, in the Coulomb case the solid curves are convex at small $n$, but all that is important for present purposes is that the two curves cross with a finite relative slope at $n=n_c$.
} \
  \label{fig:fig2}
\end{figure}
If the interactions are of intermediate range, $2 < x \le 3$ the macroscopically phase separated state still has lower free energy than the uniform density states.
(For $x \le 2$, any macroscopically inhomogeneous state has infinite free energy density.)

For present purposes, the reason interactions with  a range $ x \le 3$ differ fundamentally from shorter-range interactions is that 
the surface tension is nonlocal, and is negative if
the characteristic size of the domains, $L$, is big enough.
In other words, one can still express the free energy of a general microemulsion in terms of a double integral over the interfaces between domains, but the integral is non-local.  Specifically, any pattern of domains can be represented by a set of directed non-intersecting loops corresponding to the boundaries between the two phases, with the direction assigned so that when walking along the boundary, the WC phase is always to your left.
In Appendix \ref{thermoappend} by a slight generalization of the calculations in  Ref. \onlinecite{Reza},  we show that the resulting free energy density can then be written as
\begin{eqnarray}
F_{ME}=\tilde \mu [f_{WC}-f_{FL}]
+\Omega^{-1}\int  d{ l} \sigma_0,
-
\int d {\bf l} \cdot d {\bf l^\prime} \tilde{V}(|{\bf l}-{\bf l^\prime}|)
\label{Fmicro}
\end{eqnarray}
where $f_{WC} =1-f_{FL}$ is the areal fraction of WC, the integrals over $d {\bf l}$ are along the interfaces, and  $\tilde V(r)\sim Q^{-1} r^{-|2-x|}$. (An explicit expression for $\tilde V$ in terms of $V$ is given in Appendix \ref{thermoappend}.) 
Here $ \sigma_0= \sigma_0(\theta)$ is the contribution to the surface tension from the short-range parts of the free energy, where $\theta$ is the orientation of the interface relative to the a symmetry axes of the adjacent WC phase.  Were $\tilde{V}(r)$ shorter range than $ 1/r$ ({\it e.g.} for $x>3$), there would be a well defined local surface tension, $ \sigma^{eff} = \sigma_0-\int d {\bf  l} \tilde{V}(|{\bf l}|)$, so whether or not there are microemulsion phases would depend whether or not $\sigma^{eff}$ is negative.  However, for the Coulomb and dipolar cases (and all cases in between), the surface tension is scale dependent and negative on large distances.  Consequently, the free energy per unit length associated with long enough segments of interface is always negative, so the occurrence of microemulsion phases is universal.

It is striking that $\tilde V$ has the same power law behavior for $x=2\pm \delta x$, and in particular for the two physically important cases of $x=1$ and $x=3$.  The state which competes with the microemulsions is the uniform state for $1\le x\le 2$ and the macroscopically phase separated state for $2 < x$, but the mean-field theory which determines the size and shape of the regions of the two phases that make up the ME is the same regardless of the sign of $\delta x$.  Specifically,
in both the Coulomb and dipolar cases,  Eq. \ref{Fmicro}  can be expressed as
\be
F_{ME}=\tilde{\mu}(n-n_{c})(f_{WC}-f_{FL})+\Omega^{-1}\left\{ \int d{l}\sigma_0
 -\sigma_{1}\int  \frac{d{\bf l} \cdot d{\bf l}'}{\sqrt{|{\bf l}-{\bf l}'|^{2}+a^{2}}}\right\}
\label{Fmicro1}
\ee
where $\sigma_1$ (for which an explicit expression is given in Appendix \ref{thermoappend}) is a microscopic coupling constant with units of a surface tension.
 We have included a short-distance cutoff, $a$, in Eq. \ref{Fmicro1} which is roughly equal to the smaller
  of $d$ and $\xi$.
The chemical-potential like term, $\tilde \mu$, in Eqs. \ref{Fmicro} and \ref{Fmicro1} is a function of
 the average density and the various interaction strengths.  In the absence of the negative contribution to
  the surface tension from  the long-range interactions, $f_{WC}$ would jump from its minimal value,
   $f_{WC}=0$, to its maximal value, $f_{WC}=1$ as $\tilde \mu$ passes from positive to negative.
 The negative effective surface tension is responsible for the existence of mixed phases. The mean-field
  phase diagram is found by minimizing  Eq. \ref{Fmicro} with respect to the size and shape of the domains.

Notice, however, that $\tilde \mu$ is not precisely a chemical potential, as it is conjugate to $f_{WC}$ rather than the mean density, $\bar n$. In general, $\bar n$ as a function of
 $\tilde \mu$ can be deduced  by inverting the analysis that lead to Eq. \ref{Fmicro1}:  once the optimal solution of Eq. \ref{Fmicro1} is found, the corresponding density profile can be then determined by means of
  Eq. \ref{nprofile} in Appendix \ref{thermoappend}.
 For the Coulomb case, the density profile in each region is
non-uniform, and depends on the size and shape
 of the microemulsion.   Specifically, the density deviates from its average value most strongly at the edges of the  dipole-layer (of width $\sim \xi$) associated with the WC - FL interface, and then decays toward the mean in proportion to the reciprocal distance from the interface.
In the dipolar case, if $L$ is sufficiently large, the mean-density within each
   region is essentially that of the corresponding uniform phase,
    {\it i.e.} $\bar n \approx  f_{WC} n_{+} + f_{FL} n_{-}$.
  In this case, the values of $f_{WC}$ and $f_{FL}$  can be determined approximately 
by 
the Maxwell construction.  

One consequence of this is that, for the dipolar case is that $f_{WC}\approx (n_+ - \bar n)/(n_+-n_-)$, independent of the detailed nature of the ME, so the mean-field shape  of the ME can be found by minimizing just the surface free energy, 
second two terms in Eq. \ref{Fmicro1} at fixed $f_{WC}$, as was done in Refs. \onlinecite{SpivakPS,SpivakKivelsonPS}.
To obtain further insight into the meaning of the final non-local term, we can compare it to the standard expression\cite{LLECM} for a finite size capacitor of area $S$ and perimeter $L$:$C=S/4\pi d+(L/8 \pi^{2})\ln (\sqrt{S}/d) +{ \cal O} (d)$.  The same logarithmic term arises in integrating Eq. \ref{Fmicro1} over the boundary of an isolated domain, since indeed it has the same interpretation in terms of fringing fields.

\subsection{Electronic Microemulsions in an ``Ideal'' MOSFET}
\label{ideal}

Let us now consider explicitly  the mean-field phase diagram
of the 2DEG in the presence of a ground plane displaced by a
distance, $d$, so that $V(r) \sim e^2/ r$ for $r\ll d$ and
$V(r) \sim e^2d^2/ r^3$ for $r \gg d$. (The Coulomb case
corresponds to the limit $d\to \infty$.)  The mean-field phase
diagram for this problem is shown schematically in Fig.
\ref{fig:fig1}.

If there were only the uniform FL and WC phases, there would be a single  line of putative
first order phase transitions, indicated by the dashed line in Fig. \ref{fig:fig1}.  A peculiarity of this phase diagram is the re-entrant character of
transition as a function of $n$.  For Coulomb interactions, the ratio of the potential to kinetic energy, $r_s$, decreases with increasing density, so where $nd^2 \gg 1$, increasing density always favors the  FL.  Conversely, for short-range interactions, as in $^3$He, $r_s$ is an increasing function of density, and hence increasing density favors the WC. From this point of view, dipolar interactions are much the same as any other short-range repulsions, so where $nd^2 \leq 1$, the FL is the low density phase.
One implication of this analysis is that there exists a critical value of the ratio, $d/a_B = \alpha_c$,
 such that for $d < \alpha_c a_B$, there is no WC phase at any density.  In the same way that
the value of $r_{s}$ at which the energy of the uniform WC and FL cross estimated in \cite{cip} turned out to be large,
we think that is likely that 
that $\alpha_c\gg 1$,
but as far as we know, this issue has not been addressed by any of the numerical methods that have been applied to the problem.

The microemulsion phases exist between the two solid lines in Fig.\ref{fig:fig1}.
We will characterize them first at mean field level and later with fluctuation corrections included.  At mean field level,  we must minimize the free energy with respect to the size, shape, and density profile of the domains of the two phases.  
  A complete analysis of the patterns that minimize  Eq. \ref{Fmicro1} does not exist.  However, the general
structure of the solutions has been discussed in Refs. \cite{SpivakPS,SpivakKivelsonPS,Reza},
as summarized in Fig. \ref{fig:fig3}.

\begin{figure}
  \centerline{\epsfxsize=10cm \epsfbox{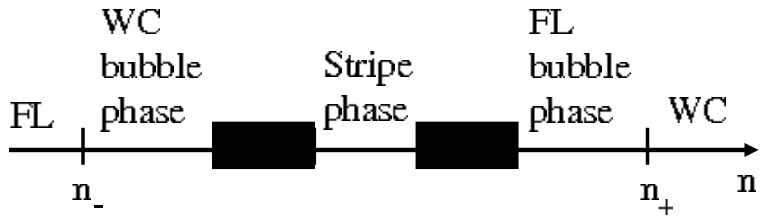}}
  \caption{A schematic picture of the mean field micro-emulsion phase diagram of an ``ideal'' MOSFET.  The solid bars indicate regions of more complicated (super-micro-emulsion) phases, consisting of a mixture of stripe and bubble crystal regions.
} \
  \label{fig:fig3}
\end{figure}

In the middle of the phase separated region, where  $f_{WC} = f_{FL}$, the free energy is always minimized by a stripe array.   The integrals in Eq. \ref{Fmicro1} are easily done  for stripes of width $L \gg a$, with the result
\be
F_{stripe}=L^{-1}[ \sigma_0 - \sigma_1 \log(L/a)].
\label{Fstripe}
\ee
This expression can be  minimized with respect to $L$ to obtain the equilibrium stripe width,
\be 
L=L_0 = a \exp[\sigma_0/\sigma_1+1]. 
\label{L0}
 \ee
So long as $L_0\gg a$ is large ({\it i.e.} if $\sigma_0 > \sigma_1$), the considerations that lead to this result are self-consistent.  If $L_0$ is comparable to the spacing between electrons, fluctuations and other effects could qualitatively affect the nature of the equilibrium state. However, even in this case it still follows from the large $L$ behavior of Eq. \ref{Fstripe} that there is a ME state that has lower energy than the uniform or macroscopically phase separated states, even if the present mean-field theory is insufficiently microscopic to describe it.

In a simple (but not generic) situation, where the surface tension is strongly anisotropic
 across entire phase diagram, there are only uniform and stripe phases  \cite{SpivakKivelsonPS}.
 (This can readily happen, for instance, in semiconductor heterostructures grown along a  cyrstalographic direction with a highly anisotropoic effective mass tensor.)
In this case, as $\bar n$ tends towards $n_-$, the stripes of FL continue to have a characteristic width  of order $L_{0}$, while the intervening regions of WC have a diverging width as $f_{WC} \to 1$.  
(The same story pertains to the limit $\bar n \to n_+$, with the role of the WC and FL regions interchanged.)
 We will refer to this type of transition (which recurs several times in our analysis) as a ``Lifshitz transition,'' despite some differences with the usual Lifshitz case.
\footnote{ A conventional Lifshitz point
is characterized by the vanishing amplitude of the order parameter
and a diverging period of the superstructure. Thus it is different
from the present case in which  the distance between bubbles or stripes diverges
at the critical point, while the size of the bubbles or stripes and the density
inside them remain constant. Hoowever, for the lack of a better name, we will nevertheless
use  ``Lifshitz point'' to describe these transitions. }

When $\sigma$ is less anisotropic,  including the usual WC case when $\sigma$ has the same 6-fold rotational symmetry as the triangular lattice, the  number of distinct phases in the mean-field phase diagram is much larger:
In the limit when the fraction of minority phases is small,
$F_{ME}$ is minimized by a bubble crystal phase. For $n$ near $n_+$, this phase consists of a triangular lattice of  crystallites
 embedded in the
liquid.  Conversely, for $n$ near $n_-$, the bubble phase consists of puddles of FL embedded in
WC. The distance between the bubbles diverges, but their radius approaches a constant, , $L_B \sim L_0$,
as the fraction
of
 the minority phase tends to zero, so again there is a ``Lifshitz transition'' to the uniform phase.  
  
 It follows  from Eq. \ref{Fmicro1} that  the interaction between  bubbles at large separation $R
 \gg L_0$ always is dipolar,
 as
\be
\label{Vbubble}
V^{(B)}(R)\sim \frac{\sigma_1 L^{4}_{B}}{ R^{3}}.
\ee

A direct transition from a bubble  to a  stripe phase at a critical value of $f_{WC}$ would, at least at mean-field level, be first order, and hence is forbidden.  Iterating the above analysis, this putative transition must give way to a new set of super-microemulsitons, consisting of regions of bubble phase and regions of stripe phase.  These phases are indicated  in Fig. \ref{fig:fig3} by black regions.  Since the surface tension of a stripe phase is highly anisotropic, it is probable that one super-stripe phase (as opposed to an infinite hierarchy of them)  result from this analysis, as has been discussed in Ref. \cite{SpivakKivelsonPS}

The various possible ME phases can be grossly classified by their broken symmetries.  They can also be classified by whether they are insulating or fluid.  Many of the phases can productively be thought of as quantum analogues of classical liquid crystals ~\cite{fradkin}:  For instance, a stripe phase which breaks translational symmetry in one direction, only, but conducts readily in the transverse direction is analogous to a smectic liquid crystal, and therefore can be referred to as an ``electron smectic.''  This analogy also permits us to conceive of phases that could be thought of as partially melted versions of the various mean-field phases:  A partially melted stripe phase, for instance, can naturally give rise to an ``electron nematic'' phase ~\cite{fradkin,fradkinqh,vadim} in which translation symmetry is restored, but the preferred stripe direction remains distinct, leading to a uniform fluid state with a spontaneously chosen nematic axis.  For the most part, we will leave the exploration of these many rich possibilities to a future study.

\subsection{Thermal physics:  the Pomeranchuk effect}
\label{Pomeranchuk}

The thermal evolution of the microemulsion phases at low
temperatures is dominated by the spin degrees of freedom
\cite{SpivakPS,SpivakKivelsonPS}. In direct analogy with the Pomeranchuk effect
familiar from the theory of $^3$He, the spin entropy of the WC
 is large compared to that of the FL, so the system tends to freeze upon heating!

In the FL state, the electron spin degrees of freedom are quenched due to the Fermi statistics.
  Consequently, the Fermi liquid entropy density is small, $S_{FL} \sim n (T/E_F)$, at  $T \ll E_F$.
    Conversely, at $T>J$ the spin entropy of the WC is large.
   The exchange interaction between localized
     electrons is exponentially small, $J \sim \exp(-\alpha \sqrt{r_s})$. (See, for a review,  ~\cite{Roger}). For example,  a  WKB calculation ~\cite{sudip98}  for $r_s=30$  yields $J\approx 5$mK.
Therefore, at essentially all experimentally accessible temperatures (as of this date), $T > J$.   Consequently $S_{WC}\sim n \ln 2 \gg S_{FL}$, 
and
\be
F_{WC}(T) \approx F_{WC}(0) - Tn\log[2].
\ee
This results in a $T$ dependent shift in the ``center'' of the microemulsion phase, $n_{c}(T)$:
\be
n_c(T) = n_c(0)[ 1 + \zeta_1^{-1}T\log(2) ],
\label{ncofT}
\ee
where $\zeta_1$ is defined in Eq. \ref{Fps}.  As a result, anywhere in the microemulsion phase, $f_{WC}$ increases linearly with $T$.
This 
behavior will dominate the $T$ dependence of many important physical properties of the system at low $T$.

At $T>E_{F}$ the liquid is not degenerate and therefore there is none of the entropy suppression characteristic of the FL. Only at these high temperatures does the liquid entropy exceed the WC entropy. One consequence of this is that for densities $n\sim n_{c}$,
the WC melting temperature $T_{m}\sim E_{F}$.
Thus there is a single temperature scale, $T\sim E_F$, that governs the crossover to a low temperature electronic microemulsion regime.
The effect of temperature on the phase diagram of electronic microemulsions is shown qualitatively in Fig. \ref{fig:fig4}.
 Of course, within the microemulsion regime there may be a number of thermal phase transitions between states with different patterns of symmetry breaking, such as stripes, bubbles, nematics, etc.  These more detailed thermal transitions will be addressed in the next susbsection.

Since the Pomeranchuk effect
is the physics of spin entropy,
 similar considerations govern the $H_{\|}$ dependence  of the phase diagram (See Fig. \ref{fig:fig5}).
A parallel magnetic field couples only to the electron spins.
Since the spin susceptibility $\chi_{WC}\gg \chi_{FL}$, the corresponding magnetization, $M_{WC}\gg M_{FL}$ at small $H_{\|}$.  Since the free energy of the system contains the term $-MH_{\|}$ there is a $H_{\|}$ induced increase of the WC fraction over a wide range of circumstances.

 For $T_m \gg T > J$, the combined $H_{\|}$ and $T$ dependence of $F_{WC}$ is  that of 
 free spins:
\ba
&&F_{WC}(T,H_{\|}) = F_{WC}(0,0) - Tn {\cal S}
\\
&&\ \  \ \ \ \ \ \ {\cal S} \equiv  \log[2\cosh(H_{\|}/T)] .\nonumber
\ea
 In this same range of temperatures, the $T$ dependence of $F_{FL}$ is negligible, and so long as $H_{\|} \ll E_F$, its $H_{\|}$ dependence can be neglected as well.  In this regime, therefore, the fraction of WC can be expressed as a scaling function
\ba
\label{pomWC}
&&f_{WC}(n,H_{\|},T) = f_{WC}(n,0,0) + g(n,T{\cal S})  \\
&& \ \ \ \ \ \ \ \ {\rm for} \ \ T  \ \ \ll E_F \ \ {\rm \&} \ \   H_{\|} \ll H^{\star}\equiv E_F,
\nonumber
\ea
where $g$ is a non-negative monotonically increasing (initially linear) function of $T{\cal S}$, and $g(n,0)=0$.
At high enough magnetic field, when all the spins are polarized, the Pomeranchuk effect is completely supressed, so that
\ba
&&f_{WC}(n,H_{\|},T) = f_{WC}(n,\infty,0)   \\
&&\ \ \ \ \ \ \ \  {\rm for} \ \ T \ \ \ll E_F  \ \ {\rm \&} \ \  H_{\|} > H^{\star}.
\nonumber
\ea

\subsubsection{Experimental consequences of the Pomeranchuk effect}
\label{experfig}
 Since the resistance of the WC and FL are very different, the  Pomeranchuk effect 
is one of the most directly testable features of the microemulsion phases. 
The expected $T$ and $H_{\|}$ dependences of the resistance of a micrio-emulsion phase $\rho$ are shown qualitatively in  Figs. 7 and 8 in Sec. \ref{comparison}. The fact that $\rho(T, H_{\|})$ increases with $T$ and $H_{\|}$ at low $T$
reflects the fact that the fraction of WC increases as $T$ and $H_{\|}$ increase. 
The fact that at $H_{\|}>H^{\star}$,  the $T$ and $H_{\|}$ dependences of $\rho(H_{\|})$ are  quenched reflects the fact that there is no spin-entropy in the fully polarized system, so $f_{WC}(T, H_{\|})$
no longer depends strongly on these quantities.
At higher $T$, the WC melts and the decrease of $\rho(T)$  reflects the corresponding $T$-dependence of the viscosity of the electron liquid.
 A more detailed analysis will be presented in Section. \ref{eta}.

\subsection{Fluctuation effects}
 \label{fluctuations}

The mean field treatment gives a reasonable 0$^{th}$ order description of the
micro-emulsion phases when the characteristic spacial scale is
large, $nL_{0}^2\gg 1$, so that the elementary structures of the micro-emulsion contain many electrons.
   In this case, where the mean-field theory fails, it can only be due to the effects of collective long-wave-length fluctuations on length scales large compared to $L_0$, rather than due to large-amplitude fluctuations on length scales of order $L_0$.

Let us begin by considering the case of the bubble crystal.
Near the  Lifshitz point, where the spacing between bubbles $R\rightarrow \infty$,
the thermal energy of bubbles is larger than the interaction energy Eq. \ref{Vbubble},
so the melting temperature of the bubble superlattice must vanish as $R\to \infty$.   On the other hand, for fixed  $R$,
the strength of the interaction between bubbles is finite, so the the bubble crystal phase survives
thermal fluctuations in the usual sense that the correlations of bubble positions exhibit power-law decay.
 The nature of the thermal transition between the bubble crystal phase and the uniform phase is not, presently, settled. Of course a direct first order phase transition is forbidden. One possibility is that there
is  a sequence of two transitions with an intermediate hexatic phase.
Alternatively, there may be a further set of hierarchical microemulsion phases.

Let us now discuss the role of the quantum fluctuations of the positions of the bubbles.
At low bubble concentration, $n_{B} \propto R^{-2}$, they can be characterized by an effective mass $M_{B}$. (See Sec. \ref{bubbleqp}.)
Thus we can introduce a  ratio between the potential energy of the interbubble interaction and the
zero-point kinetic energy  $r_s^{eff} \sim R^{-1}M_{B}$. Since $r_s^{eff}$ becomes small at small values of $n_{B}$, at $T=0$,  the bubble crystal phase is unstable  below a critical value of $n_{B}$.  Consequently, the Lifshitz point of the FL-WC bubble phase transition is preempted by quantum fluctuations.
A quantum melted WC bubble phase has the same symmetry as a liquid.
The nature of the quantum phase transition between the bubble liquid and bubble crystalline phases is currently unknown.

Let us now consider the role of thermal fluctuations on the stripe phases.
If the Hamiltonian is rotationally invariant then the smectic  phase is
unstable at any non-zero temperature to the proliferation of dislocations. Thus the mean
field smectic phase is replaced by a nematic phase\cite{nelsonpelcovitz}, which in keeping with the Mermin-Wagner
theorem, does not actually break rotational symmetry, but rather has power-law orientational order.
We refer the reader to Ref.~\cite{SpivakKivelsonPS} for details.

\subsection{Crossovers at higher temperatures}
\label{highT}

Although it is beyond the primary focus of the present paper, it is worth commenting on the nature of some of the higher temperature behavior of the electron fluid at large $r_s$.

At very low electron densities and $T=0$, the WC is classical, and there are two
characteristic energies in the system:  the electron interaction energy $V= e^{2}\sqrt{\pi n}$ and
the plasma frequency (evaluated  at wave-vectors of order $n^{1/2}$),  $\Omega_P\sim \sqrt{e^{2}n^{3/2}/m}\ll V$.
The classical melting occurs ~\cite{chakravartychester,stern} at a critical temperature $T_{cl}\approx V/125$ where $V\gg T_{cl}\gg \Omega_P$.
For $T>T_{cl}$, there are still two distinct temperature intervals:

1.  $V > T > T_{cl}$:  Here the electron fluid is a highly correlated classical fluid.  There are many examples of such fluids - indeed, most fluids fall in this regime ~\cite{Frenkel}.  In our opinion, the present theoretical understanding of such strongly correlated fluids is rudimentary, but there is a wealth of empirical information that we can draw on to make fairly confident statements about the expected behaviors.  In particular, the viscosity is widely observed ~\cite{Frenkel,DAD} to be an exponentially increasing function of decreasing temperature, $\eta \sim \exp[T_0/T]$ where $T_0\sim V$ .

2.  $T > V$:  Here the electron system forms a classical weakly interacting gas, where the viscosity is  an increasing function of $T$ because the electron mean free path is an increasing function of energy
~\cite{LifPitPhysKin}.

 \begin{figure}
  \centerline{\epsfxsize=10cm \epsfbox{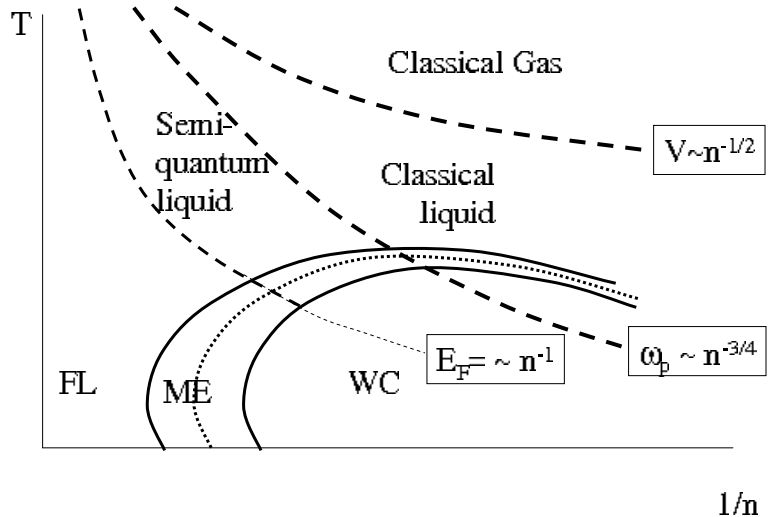}}
  \caption{The phase diagram of the 2D electron gas. FL, WC and ME stand for the Fermi liquid, Wigner Crystal and electronic micro-emulsion phases respectively. The dashed lines indicates the $n$-dependencies of the Fermi energy, the plasma frequency, and the inter-electron interaction energy.} \
  \label{fig:fig6}
\end{figure}

Let us now consider the case when the electron density is near the critical density
of the putative quantum FL-WC transition.
The  critical value of $r_s=r_s^\star$
is ``of order one''.   However, it turns out\cite{cip} to be numerically quite large: $r_s^\star\approx 40$,
so we hope to be forgiven for considering it as a large parameter.
In this case there are three characteristic energies:  1)  $E_F$, or more properly
~\cite{AndreevSQL,AndreevSQL1}, the renormalized Fermi energy, $E_F^\star < E_F$, which contains
 a renormalized mass, 2) the interaction energy  $V=r_s E_F$, and 3) the plasma frequency
  $\Omega_P\sim \sqrt{E_F V} =\sqrt{r_s} E_F$. For large $r_s$, these energies are quite distinct,
   with $V > \Omega_P >E_F^\star$, as shown schematically in Fig. \ref{fig:fig6}.
Thus, we can identify three distinct regimes of temperature, above the transition temperature to
 the microemulsion phases, where the electron fluid can be expected to exhibit different qualitative behaviors:

1)  $T < E_F^\star$:  Here, of course, as long as the system is fluid, quantum coherence is important as are the Fermi statistics.

2)  $\Omega_P > T > E_F^\star$:  Here the electron-fluid is in a semiquantum regime ~\cite{AndreevSQL,AndreevSQL1}.  Quantum coherence, and specifically the Fermi statistics of the electrons, are presumably not terribly important ~\cite{He_3He_4}.
The motion of electrons
consists mostly of quantum oscillations around certain positions,
with a frequency $\Omega_P\gg T$, which is also large compared to the rate, $1/\tau_{cage}$, at which electrons escape from the  local ``cage'' formed by the neighboring electrons, $\Omega_P \gg  1/\tau_{cage}$.
  As far as we know, there is no theory of the behavior of electronic fluids in this regime. However,  a similar regime has been discussed in the framework of
liquid $^{3}He$ and $^{4}He$ ~\cite{AndreevSQL,AndreevSQL1}, where it was argued that
$1/\tau_{cage}\sim
E_F^*$.

3)  $V > T > \Omega_P$:  Here the electron fluid is a highly correlated classical fluid.

4)  $T > V$:  Here the electron fluid forms a classical gas.

\section{Beyond Boltzman Transport in Clean, Correlated Electron Fluids}
 \label{beyond}

 This section is something of a digression.  The issue addressed here is how the resistivity of a highly correlated fluid in the presence of weak disorder is related to the hydrodynamic properties of the ideal fluid in the absence of disorder.  The considerations here are general, and apply to a FL as well as to an electronic microemulsion.  Despite this, much of the analysis has not been carried out previously, to the best of our knowledge.  So we will discuss the general problem in the present section, and then apply the results to compute the transport properties of electronic microemulsions in the following sections.

The resistivity, $\rho$, of the electron gas is determined by the electron-phonon, electron-electron (ee) and
electron-impurity (e-i) interactions.
At the low temperatures of interest,  electron-phonon scattering is negligible.
  Since $k_F$ is much smaller than the reciprocal lattice vector of the atomic lattice, e-e Umklapp scattering is also negligible which means that e-e scattering processes conserve quasi-momentum and so do not directly contribute to the resistance.
 Thus, ultimately, the resistivity is determined by the electron-impurity interactions.
  However, the actual mechanism  of momentum transfer from the electron system
 to the lattice
 can vary depending on circumstances.
 In the majority of cases considered in the literature (see for example \cite{LifPitPhysKin}) the current distribution is treated as statistically uniform and the e-e and e-i scattering are treated as uncorrelated, and are described by different parts of the scattering integral
in the Boltzmann kinetic equation. (See Subsection \ref{drude}.)  In this case, e-e scattering can contribute to the resistance only to the extent that it affects the form of the electron distribution function. In particular e-e scattering does not contribute to the resistance {\it at all } if the electron-impurity scattering rate is energy independent.

In Subsection  \ref{relation}, we show that the assumption concerning the uncorrelated nature of e-e and e-i scattering is not valid
 when the
electron electron interaction is strong and the disorder is sufficiently weak.
 The main result of this subsection is that
 in this limit, the resistance of the sample is proportional to the electron viscosity,
  \begin{equation}
\rho\propto \eta.
\label{Rhydro}
\end{equation}
To appreciate how   significantly this result differs from the usual Drude result,
note that $\eta$ in a FL is proportional to the e-e mean-free path, $\ell_{ee}$. (See Subsection \ref{eta}.)
Since $\ell_{ee}$
is a decreasing function of $T$, this implies the opposite sign of the temperature dependence of typical metals!  More fundamentally, the strong interference between correlation effects and impurity scattering implied by this relation is impossible in the context of the Boltzman equation.

We derive Eq. \ref{Rhydro} in several limiting cases in the following.  The simplest case (See Subsection \ref{pinball}.) is where there is a dilute set of large radius obstacles placed in an otherwise ideal FL.
Then the force on the obstacles from the moving liquid is given by the Stokes formula and is proportional to $\eta$.
  The important point is that, under any circumstance in which the disorder leads to a non-uniform distribution of the current on scales long compared to the appropriate microscopic healing length ($\ell_{ee}$ in the case of the FL), the resistance is more or less proportional to $\eta$, as in Eq. \ref{Rhydro},
and so will exhibit $T$ and $H_{\|}$ dependencies that reflect the corresponding dependencies of the viscosity $\eta(T, H_{\|})$ in the ideal system.

\subsection{Boltzman-Drude Resistivity}
\label{drude}

 Usually electron-electron  and electron-inpurity
scattering in semiconductors and metals
 are described by a Boltzmann kinetic equation
 \begin{equation}
 {\bf E}\cdot \nabla_{{\bf r}} n_{{\bf k}}({\bf r})=-I_{ee}({\bf k},[n_{{\bf k}}])-I_{ei}({\bf k},[n_{{\bf k}}])
\label{Boltz}
 \end{equation}
 where $I_{ee}$ and $I_{ei}$ are electron-electron and electron-impurity scattering integrals which are
  functionals of the electron distribution, $n_{{\bf k}}({\bf r})$, but are independent of ${\bf
  r}$; and ${\bf E}$ is the electric field.
 One of the assumptions which leads to Eq. \ref{Boltz} is the absence of correlations between
 the e-e and e-i scattering, which in turn is related to an assumption that
 in a statistically uniform system the distribution function of the electrons $n_{{\bf k}}$ is
 spatially uniform, {\it i.e.} it is the same near impurities and between them.

Since the e-e scattering  conserves the total quasi-momentum of the electrons,
 only electron-impurity scattering contributes to the resistivity leading to the Drude formula
 (See for example \cite{Abrikosov})
 \begin{equation}
 \rho_{DR} = \frac{3m}{e^{2}\nu v_{F}^{2}\tau_{i}}
 \end{equation}
 where
 $\nu$ is the density of states at the Fermi level, $v_{F}$ is the Fermi velocity,
 $1/\tau_{i}$ is the electron-impurity transport scattering rate computed in the usual way.

If the impurity potential is treated in the white noise approximation, $\tau_{i}$ is independent of the electron Fermi wave length.
 Thus $\rho(T, H_{\|})$ is independent of $T$ and $H_{\|}$.
Typically, though, the electron-impurity cross-section is a
decreasing function of the energy so $\rho(T, H_{\|})$ decreases
both with $T$ and with $H_{\|}$.

 As is well known, the Drude-Boltzman description breaks down at low enough $T$, due to quantum interference effects ;  ``weak localization'' phenomena is an example of this sort of corrections.  However, in the weak disorder limit, these corrections are only of large magnitude at exponentially low temperatures.

What we discuss below is a different sort of breakdown of the Boltzman description, which is important in the limit of small disorder,  especially in the case of
 low-dimensional highly correlated systems.
 Specifically, the assumption of Boltzman theory that the current distribution is uniform  is valid only so long as the impurities responsible for the resistivity are sufficiently dense that they  can be treated as self-averaging.  Roughly, this requires that electron-impurity mean free path is smaller than electron-electron one $\ell_{ei} \ll \ell_{ee}$.

\subsection{Relation between resistance and viscosity}
 \label{relation}

For sufficiently weak disorder, the motion of a FL at length scales larger than $\ell_{ee}$ can be treated
 in the context of the hydrodynamic  Navier-Stokes equations \cite{LandLifHydr}
\begin{equation}
n\left [\frac{\partial U_{i}}{\partial t}+ U_i\frac{\partial
U_j}{\partial r_j} + \frac{\partial F}{\partial r_{i}}\right ]  +
\frac {\partial \Pi_{ij}} {\partial r_j}=- n \frac{d \Phi({\bf
r})}{d r_{i}}
  \label{navier}
\end{equation}
where   $U_{i}({\bf r})$ are the components of the electron
hydrodynamic velocity, $F$ is the free energy density, $\Pi_{ij}$ is
the momentum current tensor,
 and $\Phi({\bf r})$ is an external potential ({\it i.e.} the impurity potential).  $\Pi_{ij}$,
  in turn, can be expressed as
\begin{equation}
\Pi_{ik}=\eta_{1}\left[\frac {d U_{i}}{d r_{k}}+\frac {d U_{k}}{d
r_{i}}-\frac{2}{3}\delta_{ik}\frac {d U_{l}}{d r_{l}} \right ]
+\eta_{2}\delta_{ik}\frac {d U_{l}}{d r_{l}}.
 \label{navier2}
\end{equation}
Here  $\eta_{1,2}$  are the first and the second
viscosities respectively.
The first viscosity $\eta_{1}$ determines the viscous drag on an object moving through the fluid, and so is the more important one for present purposes.   The second viscosity $\eta_{2}$ is associated with the rate of equilibration of density variations.

\subsubsection{Dilute large obstacles}
\label{pinball}

The first case we will consider is one in which the ``disorder'' consists of a dilute concentration, $N_i$ of large obstacles (radius $\lambda\gg \ell_{ee}$ but still $N_i\lambda^2 \ll 1$) in the electron liquid. (For example, these obstacles could be pinned bubbles of WC, as we will discuss in Subsection \ref{bubble}B, below.)  The resistance can be computed straightforwardly by equating the electric field to the force exerted on a flowing liquid by a stationary object, which is given by the  Stokes formula. As a result we have an expression for the resistivity
 \begin{equation}
 \rho= N_{i}\eta \frac{1}{|\ln(\lambda N_{i}^{1/2})|},
  \label{stokes}
 \end{equation}
  The fact that $\rho$ depends  only logarithmically on $\lambda$ and  is a slightly superlinear function of
   $N_i$ reflects the  ``Stokes paradox'' in 2D hydrodynamics \cite{LandLifHydr}.

\subsubsection{Dilute microscopic impurities}

The case of a 2D FL with $\ell_{ee}$ larger than the electron-impurity
cross-section, but still small compared to the spacing between impurities was analyzed in Ref. \cite{HruskaSpivak}.  The result is:
\begin{equation}
\rho= \frac {\rho_{DR}}{[1+\frac{\alpha \lambda v_{F}}{\eta}\ln(l_{ee}N_{i}^{1/2})]}
 \label{hurska}
 \end{equation}
 where  $\alpha$ is a number of order 1.
 The logarithm in the
 denominator of this formula is also a consequence of the Stokes
 paradox in 2D.  Note that when the logarithmic term is large, this again gives a resistivity that
  is proportional to the viscosity.

  \subsubsection{Weak, smooth disorder potential}

 Another case which can be analyzed readily is that in which the scattering potential is smooth and is characterized a length scale of $\lambda\gg l_{ee}$ and of small amplitude $V_{0}$.
\begin{equation}
\rho\sim \eta (\frac{V_{0}}{E_{F}})^{2}\frac{1}{\lambda^{3}}
\end{equation}

\subsection{Hydrodynamics of an ``ideal'' electron liquid}
 \label{eta}

 Before embarking (in the following sections) on the analysis for the 
 micro-emulsion phases, in this section we review briefly what is known about the viscosity of FL's.
We also discuss the viscosity of strongly correlated electron liquids above the degeneracy temperature.
 Surprisingly, there are still many regimes in which the answers are not known, either theoretically or
  experimentally.
We therefore also invoke experimental results on model liquids,
especially $^{3}$He, and in one case
 even $^{4}$He, to support a phenomenological conjecture concerning the $T$ dependence of the viscosity.

\subsubsection{Viscosity of a pure Fermi liquid}

In the  Fermi liquid regime ($T\ll E_{F}$) the two viscosities are
of the same order $(\eta_{FL}\sim \eta_{1} \sim \eta_{2}$ and
\begin{equation}
\eta_{FL}   \sim np_{F}v_{F}\tau_{ee}
 \label{etaFL}
\end{equation}
is inversely proportional to the square of the temperature and
becomes very big at small $T$ \cite{LifPitPhysKin}, because
the electron electron scattering rate in the Fermi liquid
\begin{equation}
\tau^{-1}_{ee}\sim C\frac{T^{2}}{E_{F}} \ \ {\rm for} \ \ T \ll E_F
 \label{tauee}
\end{equation}
decreases with $T$. Here $C=C_{\uparrow \uparrow}+C_{\uparrow
\downarrow}$ is a coefficient of order one which is proportional
to the sum  of scattering probabilities for parallel and
anti-parallel spins.

To characterize the $H_{\|}$ dependence of the viscosity we discuss
the ratio
\begin{equation}
\frac{\eta_{FL}(0)}{\eta_{FL}(H_{\|})}=\frac{(\tau_{ee}/m^{*})_{H_{\|}=0}}
{(2\tau_{ee}/m^{*})_{H_{\|}}}
\sim \frac{(C_{\uparrow \uparrow}m^{*})_{H_{\|}}}{4(C m^{*})_{H_{\|}=0}}
\end{equation}
 evaluated for $H_{\|} > H^\star$.
At high electron concentration when the interactions can be treated peturbatively ($r_{s}\ll 1$)  the value
 $(\eta(0)/\eta(H_{\|}>H^{*})<1$: both the
scattering amplitude and the electron effective mass decrease with increasing spin polarization of
the electron system, and the Fermi energy
increases. In the case $r_{s}>1$ the situation is more complicated
and currently it is unknown whether $m^{*}(H_{\|})$ increases or
decreases with $H_{\|}$;  the Fermi energy, however, clearly still
 increases with $H_{\|}$.
 A similar situation (large $r_s$ Fermi liquid) pertains to $^{3}He$, but the dependence of $\eta_{FL}$ on spin polarization is unknown there, as well.

\subsubsection{Viscosity of a nondegenerate strongly correlated electron liquid}

Eqs. \ref{etaFL} and \ref{tauee} are valid in a FL at any $r_{s}$ as long as the temperature is low
enough and the system is in a liquid state.  However, if $r_{s} > 1$, then (as mentioned above) there is a temperature
interval $T_{F}\ll T \ll \Omega_P=\sqrt{E_FV}$ where the fluid is still quantum but not
degenerate, and is still strongly correlated.
 The temperature
dependence of the viscosity of the electronic liquid in this interval is unknown.
 However, a similar regime exists in another strongly correlated
quantum liquid: $^{3}$He.
We know of only one theoretical paper \cite{AndreevSQL1} on the subject where
the arguments were presented that
\begin{equation}
 \eta\sim \frac{1}{T}  \ \ {\rm for }
  \ \ E_F < T < \sqrt{E_FV}\ .
  \label{andreev}
\end{equation}
Since the theoretical arguments made in Ref. \cite{AndreevSQL1} are quite general
it is plausible that
 the $T$-dependence of the electronic viscosity in the semiquantum regime
can be described by the same Eq. \ref{andreev} . Surprisingly, even experimental data to test this conjecture in $^3He$ is unavailable.
In support of it, the author of Ref. \cite{AndreevSQL1} observed that the viscosity of $^4He$ in the same temperature range is roughly consistent with a $1/T$ dependence;  since in this range of temperatures the statistics of the particles is likely unimportant, this evidence should be taken as moderately compelling.
  This means
that, as in  classical liquids, the viscosity of the
semiquantum liquid decreases with increasing $T$. The difference, however,
is that it decreases only inversely in proportion  to the
temperature, whereas the viscosity of classical correlated liquids decreases
 exponentially with $T$ ~\cite{Frenkel,DAD}.
 Presumably, the same is true for the electron-fluid
\be
\label{classicaleta}
\eta \sim \exp[-AV/T]   \ \ {\rm for }
  \ \ \sqrt{E_FV}
 < T <  V\ ,
  \ee
  where $A$ is a number of order 1.

  Lastly, for $T > V$ 
  we have is a classical
  electron  gas where the viscosity increases with $T$.

\section{The non-degenerate bubble liquid}
\label{bubble}

 The liquid phase consisting of a fluid of WC bubbles in a sea of FL is the simplest of the microemulsion fluid phases.  We will start with the simplest sub-case in which
the system is near the mean field Lifshitz point, where the bubbles are
weakly interacting, and where (as we showed in Section \ref{fluctuations}) even small thermal and/or quantum fluctuations destroy the bubble-crystalline order.
We moreover confine ourselves to the  limit in which the bubble-size is large compared to the spacing between electrons, $nL_B^2 \gg 1$, and consequently the bubbles are well defined as droplets of WC, and are relatively heavy so that quantum coherent motion of the bubbles occurs only at very low $T$.
 Slow and long
wave-length motion of the electron liquid can be described with the
help of the Navier-Stokes Eqs. \ref{navier} and \ref{navier2}. However, the viscosity of the system is very different from
the viscosity of the FL.  In particular, the viscosity depends on the
bubble concentration, $n_B$.
Thus (following Eq. \ref{Rhydro}) the resistivity in the presence of small disorder directly reflects the $T$ and $H_{\|}$ dependencies of $f_{WC}$ discussed in Sec. \ref{thermo}, above.

Because $\eta$ depends on $T$ directly, and implicitly through the $T$ dependence of $n_B$, the final result is complex and there are many special cases, as we shall see.  One important result is that, in contrast to the FL case,
the viscosity extrapolates to a finite value at $T\rightarrow 0$. Often, $\eta$ is a non-monotonic function of $T$.  Near the boundary of the bubble liquid phase, where $n_B$ is large, $\eta$ typically depends exponentially on $T$.

  \subsection{The hydrodynamics of the ideal bubble liquid phase.}

  A number of scattering processes  contribute to the hydrodynamics.  Within the majority FL portion, the same electron-electron scattering process, $\tau_{ee}$, that determines the viscosity of the uniform FL plays a major role in the bubble liquid, as well.  However, there is also an important effect of the scattering of electrons (quasi-particles) of the FL off the WC bubbles, which occurs with rate
\begin{equation}
1/\tau_{eB}\sim v_{F}\ n_{B}L_{B},
\end{equation}
where $L_B\sim L_0$ (defined in Eq. \ref{L0}) is the bubble diameter.
Although the bubbles are dilute, they are strong scatterers.

\subsubsection{ First viscosity of the dilute bubble liquid:}

 The effective scattering rate which determines the viscosity is the sum of contributions of different processes,
 \begin{equation}
\frac{1}{\tau_{eff}}=\frac{1}{\tau_{ee}}+\frac{1}{\tau_{eB}}.
 \label{taueff}
\end{equation}
 At low
temperatures $\tau_{ee}\gg \tau_{eB}$,   the
electron-bubble scattering is the dominant scattering process.
Consequently,
\begin{equation}
\eta_{1}\sim nk_{F}\frac{1}{n_{B}L_{B}},
 \label{eta1dilute}
\end{equation}
{\it i.e. the first viscosity approaches a $T$ independent constant}. This is
very different from Fermi liquid temperature dependence Eq. \ref{etaFL} according to which $\eta$
diverges at $T\rightarrow 0$.

 At higher temperatures, where $\ell_{ee}$ is less than the distance between bubbles, the viscosity is more or less dominated by the electron-electron scattering, as implied by Eq. \ref{taueff},
although with some quantitative corrections due to the presence of bubbles.  For instance, at high enough temperatures that $l_{ee}\ll L_{B}$,
 the hydrodynamic description works even on distances smaller
 than the inter-bubble distance. In this case  the viscosity   follows the
usual formula for viscosity of classical suspensions \cite{LandLifHydr}
\begin{equation}
\eta_{1}=\eta_{FL}(1+n_{B}L^{2}_{B})
\label{eta2dilute}
\end{equation}
(still assuming that $n_{B}L^{2}_{B}\ll 1$).

 The above analysis ceases to be valid at very low
temperatures where the
 motion of the gas of bubbles itself becomes quantum coherent, 
as discussed in Section \ref{quantumbubble}, below.  Nevertheless, for sufficiently dilute bubbles, the crossover to a quantum coherent
bubble fluid occurs at such low temperatures that it is not worth
belaboring.

\subsubsection {First viscosity of a dense bubble liquid}

Let us consider now the interval of temperatures and densities where the system
of bubbles is reasonably dense and close to forming a bubble crystal.
 The situation here is similar to that in the theory of classical liquids: the complete theory is not available, but it is plausable that  the flow is associated with local rearrangements of the bubble configurations.
 Thus, in this regime, as discussed in Section \ref{beyond},
\begin{equation}
\eta_{1} \sim \exp[ V_{B}/T ]
\label{etadense}
\end{equation}
where $V_{B}$ is a characteristic bubble-bubble interaction energy.

\subsubsection{Magnetic field dependence of the first viscosity}
 At low
temperatures, when Eqs. \ref{eta1dilute} and \ref{eta2dilute} are valid, the $H_{\|}$-dependence
 of the viscosity is
 determined primarily by the corresponding $H_{\|}$ dependence of $n_B$.
 Since
 $f_{WC}(H_{\|})=n_{B}(H_{\|}) L_{B}^{2}$, and  $L_B$ is only weakly $H_{\|}$ dependent,  $f_{WC}$
increases linearly with
$H_{\|}$ at small $H_{\|}$, and saturates for $H_{\|} > H^\star$.

In the case of large bubble concentrations, the bubbles
form a strongly correlated suspension and Eq. \ref{etadense} holds.
Since $n_B$ increases with increasing
$H_{\|}$, presumably $E_0$ does as well, so the viscosity is an exponentially increasing function
of $H_{\|}$.

\subsubsection{Second viscosity of the bubble liquid}

It is well known that in the presence of a long internal
relaxation time in the liquid, the second viscosity can be
much larger than the first one \cite{LandLifHydr}.
In the case of the WC bubble phase the origin of a long relaxation
time is the following:  As the total electron density changes, the
 equilibrium bubble
concentration, $n_{B}$, changes as well.  On the other hand, the nucleation of a new
bubble occurs at an extremely small rate, with an activation energy (for thermal nucleation) or a tunnel barrier (for quantum nucleation) which
is grows with increasing  $L_B$.
For example, the height of the thermal nucleation
barrier can be estimated from Eq. \ref{Fmicro}, to be
\begin{equation}
E_{0}\sim \sigma L_{B},
\end{equation}
leading to an exponentially long relaxation time,
\begin{equation}
\eta_2\sim \exp(\frac{E_{0}}{T})
\end{equation}

\subsection{Resistivity of
 the WC bubble liquid with small disorder}

In the
bubble phase there are two types of the current carriers:
electrons and bubbles.  The relaxational dynamics depends on  the interactions of the electrons (quasiparticles) and bubbles with each other and with the impurities.
Depending on
the relative strength of these different interactions, there can be many distinct regimes of behavior.
Here we consider only two.

\subsubsection*{Case 1:  Electron mobility  $\gg$
 bubble mobility}
 \label{case1}

If the bubbles are big and massive, then under many circumstances their mobility will be much smaller
than the mobility of the quasiparticles in the FL.  In this case, as far as
the  calculation of the resistivity is concerned, the bubbles
 can be treated as stationary objects which scatter the electrons.
 They differ from ordinary impurities in that their concentration
varies as a function of $T$ and $H_{\|}$, and in that they can move.

At low temperature, where $\ell_{ee}$ is long compared to the elastic mean free path,
the resistance is given by the usual
Drude formula $\rho\sim 1/\tau_{eff}$ where
\begin{equation}
\frac{1}{\tau_{eff}}=\frac{1}{\tau_{i}}+\frac{1}{\tau_{B}},
\end{equation}
and $1/\tau_B \propto n_B$.
Thus the $T$ and $H_{\|}$ dependencies of the
 resistance are determined by the
Pomeranchuk effect, as in Eqs. \ref{pomWC}.  We think that this
is the case which is most realistically realized in currently
available experiments. The characteristic $T$ and $H_{\|}$
dependencies of the
 resistance are shown in Figs.7, and 8 The saturation of the resistance at
 large $H_{\|}$ is due to the complete polarization of the electron
 liquid. The downturn of the $T$ dependence of the resistance at
 $T>E_{F}$ takes place in the region where the Wigner crystal melts
 and the temperature dependence of the resistivity $R(T)$ is determined by the temperature dependence of the viscosity,  Eq. \ref{classicaleta}.

It is worth noting that the enhancement of the resistance by the scattering from bubbles is particularly
 dramatic at low temperatures in the case where the impurity potential is smooth.  As is well known,
  a smooth potential causes little back-scattering.  However, when the bubbles are large and have
   lower than average electron density, they will tend to get stuck on the mountaintops of the
    smooth potential.   The bubbles are effective large angle scatterers.

\subsubsection*{ Case 2:  Electron mobility comparable to bubble mobility}

In the opposite limit of clean samples and at finite
 temperature, the equilibration between
the motion of the bubbles and the electrons takes place more quickly than the
momentum transfer from the electron system to the impurities. In
this case the resistance is determined by the Stokes hydrodynamic
formula and the $T$ and $H_{\|}$ dependences are expressed in
terms of the viscosity of the suspension, as in Section \ref{Pomeranchuk}.

\section{Quantum bubbles.}
\label{quantumbubble}

In this section we will discuss some of the issues concerning the role of the zero temperature
quantum fluctuations of bubbles of a minority phase 
which are beyond the treatment of the previous section.
Quantum motion of the bubbles is most important when the bubbles are small, $nL_B^2 \sim 1$ -- the situation which applies, for example, in the shaded portion of the phase diagram in Fig. \ref{fig:fig1}.
However, we do not have the analytic tools available for treating this most interesting regime of parameters. 

When the bubbles are large, $nL_B^2 \gg 1$, quantum fluctuations are small under most circumstances, and consequently the ground-state is an ordered, bubble crystal.  However, near the Lifshitz point, where the density of bubbles is small, $n_BL_B^2 \ll 1$,
quantum fluctuations are sufficient to quantum melt the bubble crystal, even when the bubbles themselves are large.
We therefore focus on this rather extreme regime as being a more tractable limit of the more general problem.

In the Subsection \ref{bubbleqp} we discuss estimates for the effective mass $M_{B}$ of WC bubbles.
In Subsection \ref{bubbleloc} we show that as long as the resistance of the sample is smaller than $h/e^{2}$, the WC bubbles are not localized by frozen disorder inspite the fact that they can be heavy objects.

\subsection{Bubbles as quasiparticles}
\label{bubbleqp}

 As discussed in Sec. \ref{fluctuations}, at low density $n_{B}$ the WC bubbles
form a quantum fluid.
  A WC bubble has a finite lifetime with respect to quantum melting, which, for a large bubble, is exponentially long.  Thus, for large WC bubbles, there is a broad intermediate time interval in which they behave as propagating quasiparticles  with a well defined effective mass $M_{B}$ and a charge
$e^\star=eM_{B}/m$. (The latter estimate follows from the fact that in a liquid, the momentum and the
flux of mass have the same form. )

There are two  different estimates of the bubble effective mass one can make, depending on the quantum character of the interface between the WC and the FL:
\begin{equation}
  \frac{M_{B}}{m}\propto \left\{
\begin{array}{ll}
  L_{B}^{2}n & \mbox{quantum smooth surface}\\
 L_{B}^{2} \Delta n  & \mbox{ quantum rough surface}
  \end{array} \right. ,
\label{MB}
\end{equation}

If the surface is quantum smooth, ( which means that in the ground state there is zero density of steps on the surface)  then the bubble moves as a
solid object  and $M_{B}$ is determined by a redistribution of the
mass of the liquid over distances of order the radius $L_{B}$ (upper formula in Eq. \ref{MB}).
On the other hand, if in the ground state the surface has steps (it is quantum rough \cite{AndreevParshin}) then
the bulk of the crystal does not, actually, move.  Rather, a strip near the rear quantum melts, and a stripe near the front coherently crystalizes.   Only an amount of mass proportional to the difference in the densities of the WC and the FL is transported as the crystal translates by this process.  Under this assumption,  we arrive at the estimate given by the lower expression in Eq. \ref{MB}.

  Now we would like to make few comments about the bubble liquid phase that occurs near the microemulsion to WC phase boundary.  Here, 
  the WC is the majority phase, and well ordered, but with dilute bubbles of FL inside it.
FL bubbles embedded in the WC are topological
objects, so the number of additional electrons associated with
each is an integer. Thus the statistics can be either fermionic or
bosonic depending on the parameters of the bubbles.

We would also like to mention the role of electron
spins in this picture: An individual bubble can be treated as a
point-like object moving through a sea of spins.  Since the exchange interaction, $J$,
  in the WC is so small, this is analogous the Nagaoka problem in that there is a strong tendency of to align the  spins in a region (a ferromagnetic polaron) around each bubble so as to minimize the bubble zero-point kinetic energy.  At $T=0$, the size of this polaron\cite{ekl} is
  \begin{equation}
  L_P \sim [n M_{B} J/\hbar^2]^{-1/4}
  \label{LP}
  \end{equation}
  where $M_B$ is the  bare effective mass of the bubble (given by an expression similar to  that in Eq. \ref{MB}). Eq. \ref{LP} has been derived under an assumption that $L_P \gg
  L_B$, or that the exchange interaction $J$ is small enough.  In this  limit, the bubble effective mass is strongly renormalized, most probably \cite{assa} by an exponential factor $M_P \to M_P \exp[+ A nL_P^2]$.

However, if $L_P$ is large, even at rather small bubble concentration the polaronic clouds begin to overlap, implying that there should be a transition to a globally ferromagnetic state.
As in the Nagaoka problem, the characteristic energy (temperature) scale of the ferromagnetism can be large compared to $J$.  

\subsection{Delocalization of WC bubbles in the presence of
 slight disorder.}
 \label{bubbleloc}

If  $\rho \ll h/e^{2}$, the weak localization corrections to the conductance are negligible on practical length and energy scales.  However,
WC bubbles are bigger and heavier than electrons. The question arises whether they are localized by weak
forzen disorder.
The purpose of this subsection is narrow:   we will show that for small  enough  $\rho$,
the WC bubbles are not localized by weak disorder. The reason for this is that the WC bubble 
can quantum melt and re-cryslallise.

To a first approximation, the bubbles
are pinned at favorable sites of
the disorder potential, which we label with a discrete index, $j$.
Since the Fermi liquid degrees of freedom are ``fast,'' we can
integrate them out to obtain an effective hamiltonian for the
bubbles.  In this case a relevant effective bubble Hamiltonian to consider has a form
\footnote{Of course, strictly speaking we should be dealing with an effective action, rather than an effective Hamiltonian;  the gapless character of the Fermi liquid means that at any energy scale, there are still lower energy excitations of the Fermi liquid which means there is  a dissipative component in the effective action.    This subtlety has negligible effects on the present results, as we will show when we return to this issue, below.}
\begin{equation}
H^{eff}= \sum_{i} \epsilon_{i} a_{i}a^{\dagger}_{i}
+ \sum_{i,j} V^{(B)}_{i,j} a^{\dagger}_{i}a_{i}a^{\dagger}_{j}a_{j} + \sum_{i\ne j} t_{i,j} \left[
a^{\dagger}_{i}a_{j}+ {\rm H.C.} \right] + \ldots
 \label{bubbleH}
\end{equation}
Here, $a^{\dagger} _j$ creates a bubble at position ${\bf R}_j$.  Since by assumption the disorder is weak and the bubbles are large, the optimal bubble at each site are all roughly the same size, $L_B$; the major effect of disorder is to pin their center of mass motion.
We assume that the energies
$\epsilon_{i}$ are randomly distributed in some interval of energies $\epsilon_{i} \in [-W,W]$.
According to Eq. \ref{Vbubble}, at long distances, $V^{(B)}_{i\neq j} \sim |{\bf R}_i-{\bf R}_j|^{-3}$.
In the mean-field groundstate, the sites with $\epsilon_{i}
<\bar{\epsilon}$ are occupied by a single bubble (The repulsion between two bubbles on the same site is effectively infinite.),
 while the sites with $\epsilon_{i}>\bar{\epsilon}$
are ``empty", which means that the system at these sites is in a liquid state.

The last term in Eq. \ref{bubbleH} which gives the bubbles quantumdynamics describes processes of transfer of bubbles from an ``occupied"  state at site ``i'' to
an ``empty''  state at ``j'', or , in other words, processes where WC bubble melts at the site $i$ and coherently recrystallizes at site "j". While the WC itself does not need to propagate long distances in this process, the density deficit associated with the bubble still needs to quantum diffuse through the FL.  As a result, as shown in Appendix \ref{delocalization} using the methods of Ref. \cite{LevitovShytov},
 as $|{\bf R}_{i}-{\bf R}_{j}| \to \infty$,
\begin{equation}
t_{i,j}\sim \frac{B}{|{\bf  R}_i - {\bf  R}_j|^{\beta}}
\label{tij}
\end{equation}
 where $\beta \sim ( \Delta n L_{B})^{2}(\rho  e^{2}/h)$ is a constant
which vanishes as $\rho\to 0$, and $B$ is an exponentially small but non-vanishing  coefficient. 

The
crucial feature of this problem, which distinguishes it from the analogous problem of Anderson localization of electrons, is that the hopping matrix $t_{i,j}$ is  long ranged.
 It is well known
\cite{Levitov} that for $\beta < D$, 
Anderson localization does not occur.
One can see this from the following:  Let us consider an interval of energy
$|\epsilon_{i}-\bar{\epsilon}|< \delta \ll W$. A typical distance between sites with energies  in this interval is
$R_{\delta}\sim (\delta \nu)^{-1/D}$, where $\nu$ is a density of states per unit energy.
Thus the ratio of the typical hopping matrix element to the energy difference between two such sites
\begin{equation}
t_{ij} /{\delta}\sim
{\delta^{-1+\beta/D}}
\end{equation}
diverges as $\delta \rightarrow 0$, provided $\beta > D$.
Thus, if $\rho$ is small enough, the bubbles are always be able to tunnel between nearly degenerate sites and hence are delocalized.

The conclusion that the bubbles are delocalized for small enough $\rho$ is apparently unavoidable.  Taking the Hamiltonian in Eqs. \ref{bubbleH}  and \ref{tij} at face value, the present analysis suggests  the existence of a bubble localization-delocalization transition at $\rho\sim h/e^{2} \Delta n L_{B}^{2}$.  However, more  complicated processes, involving the coherent motion of many bubbles, could affect this conclusion.
Finally we would like to mention that our calculations are controlled only
in the limit when $\Delta n L_{B}^{2}\gg 1$, where the amplitude $B$ in Eq. \ref{tij} is exponentially small (See Eq. \ref{B}), and consequently  the mobility of the bubbles at $T=0$ is correspondingly small.

\section{Hydrodynamics of more complex micro-emulsion phases}
\label{complex}

 In the preceding sections, we have focussed primarily on electronic microemulsion liquid phases with no spontaneously broken symmetries. At low temperatures, the same physics leads to a variety of phases with different patterns of symmetry breaking.  Because these phases are made out of clusters of electrons, rather than out of point-like electron quasiparticles, they are more analogous to the phases of complex classical fluids than to simple liquids.
 
 Microemulsion phases with different symmetries have different hydrodynamics.
  In this section, we will discuss
a few simple aspects of the problem.
Hydrodynamics deals with a set of ``slow'' variables, whose dynamics are slower than all microscopic relaxation rates.  It can be used in an asymptotic sense since the variations of hydrodynamic variables are arbitrarily slow in the long wave-length limit.  Examples of such hydrodynamic quantities are the temperature, average velocity, density, and the director in a nematic or smectic liquid crystal. The resulting sequence of hydrodynamic equations are
the conservation laws plus the linear response equations connecting the fluxes with
gradients of the parameters. These equations are asymptotically exact at times
long compared to  the characteristic relaxation time.
However the term hydrodynamics can also be used in a slightly broader sense in systems with a substantial separation of time scales to include the dynamics of any slow variables on intermediate time scales.
  Below we will use the term hydrodynamics in both senses.  We merely touch on the subject, to illustrate the richness of the possibilities.

\subsubsection{ ``Supersolid'' hydrodynamics}

Let us consider a case in which a dilute liquid of FL droplets is dissolved in a WC host.
This state has an unusual combination of liquid-like and solid-like properties.
  On the one hand, the system is characterized by  a finite shear modulus.  On the other hand, because of existence of the FL droplets, the state can flow, even in the presence of  dilute pinning centers.
To make the discussion as simple as possible, we will consider the case in which the electron spin degrees of freedom can be ignored;  this pertains  when the electrons are fully polarized by a magnetic field parallel to the film.
The remaining slow variables are related to the density and velocity of thermally excited plasmons in the WC, the bubble velocity and density, and the elastic deformations of the WC, itself.
The dynamics of this system is characterized by bubble-bubble $\tau_{bb}$, plasmon-plasmon $\tau_{pl,pl}$ and bubble-plasmon $\tau_{ b,pl}$ scattering mean free times.
Among them, a special role is played by
 umklapp-like processes  in which a momentum equal to $\hbar$ times the WC reciprocal lattice vector is transfered from the of the bubble or phonon fluid to the WC lattice.

At times longer than the umklap mean free time $\tau_{U}$, this problem is similar to any other crystal with a finite concentration of vacancies or interstitials: the hydrodynamic modes consist of underdamped longitudinal and transverse plasmon modes (where in the Coulomb and dipolar cases, the longitudinal modes have a $\sqrt{k}$ and  $k$ dispersions respectively) and a diffusive density mode. Here $k$ is the wave vector of the plasmon.
At sufficiently low temperatures,
umklap processes become ineffective. In this case, on intermediate times $\tau_{U}\gg t\gg \tau_{pl,phl},\tau_{pl,b}, \tau_{bb}$ scattering processes conserve the total quasi-momentum in the system of phonons+bubbles. In this case the hydrodynamic equations have the same form as in the case of conventional crystals where $\tau_{U}\gg \tau_{pl,pl}$ \cite{Gurevich}.  Among other things, this can lead to interesting phenomena resembling second sound in the system of bubbles+phonons.

 At low enough temperatures, when the bubble scattering rate is smaller than their energy, bubble motion on some distances has a quantum coherent character. However, as long as the bubbles are not degenerate, the quantum nature of their motion manifests itself only in the values of the transport
coefficients without changing of the form of the hydrodynamics.

At even lower temperatures, new behavior emerges which depends on the bubble statistics.
For instance, if the bubbles are bosonic, they bose-condense at low temperature.
 In this case the system of hydrodynamic equations is equivalent to the supersolid equations \cite{andreevSupersolid}.
 If the bubbles are fermionic, the system is analogous to a CDW in which a portion of the Fermi surface remains ungapped.
One of the most important consequences of the existence of a bubble fluid is that, even  at $T=0$, the system can flow in the presence of
obstacles and, therefore, it can conduct. This is very different from the case of a conventional WC, which is pinned by arbitrary small disorder.

\subsubsection{Crystallization waves in the stripe phase.}

To illustrate a different aspect of
the hydrodynamics of microemulsion phases, we consider
a stripe phase. Since the symmetry of the stripe phase is the same as that of a smectic
liquid crystal, at long enough times, the hydrodynamic equations should be identical
to the equations describing the hydrodynamics of a smectic.

On the other hand, on intermediate times, a reduced description of the system is also possible.
One has to solve the Navier-Stokes equations for a FL and the elasticity theory
for WC stripes and supplement these equations with boundary conditions at
the WC-FL interface. The nature of these conditions at $T=0$ (whether the surface is
quantum rough or quantum smooth) has been discussed in the context of the theory
of $^{4}He$ \cite{AndreevParshin}. In the  quantum rough case,  in the ground state there are steps
on the boundary and it moves in response to the infinitely small jump in chemical potential between the FL and WC. In the quantum smooth case,  there are no steps at $T=0$,
so the boundary cannot move until the difference in the chemical potentials
reaches a threshold value.
At finite temperature there are always steps on the surface.  Therefore, it can move, and there is no sharp distinction between
the quantum rough and quantum smooth cases.

The long wave length dynamics of the system
turn out to be independent of the character of the interface, including the concentration of the steps on the surface \cite{AndreevParshin}.
The reason is that the densities of the crystal and the liquid are different and as the surface moves,
the additional mass is released into the liquid. Thus the effective mass associated with the long wave length motion of the surface is determined by the mass of the moving liquid rather than by the
contribution associated with the motion of the surface itself.

More generally, the fact that the two different components of the electronic microemulsion can inter-convert leads to qualitatively new modes, not present in classical liquid crystals.
We illustrate this by considering, for the stripe phase,  the simple situation when  the crystallization
wave propagates perpendicular to the stripes. Let us introduce
the deviation of the position of the WC-FL boundary from its
equilibrium position $h(t)$. Then, according to Eq. \ref{Fmicro1}  the
potential energy per unit length of stripe
is $V^{(S)}_{pot}\sim \sigma h^{2}f_{WC}/L_{0}^{2}$.  The kinetic energy is  associated with the uniform flow of the liquid between two
stripes.  The flux density of mass released into the liquid by the
motion of a boundary is $J=m\Delta n \dot{h}(t)=mn U$,
where $U$ is the velocity of the liquid. As a result, at $f_{WC}\ll 1$
the kinetic energy of FL is $=m n L_{0} f_{WC}^{-1}(\Delta
n)^{2}\dot{h}^{2}(t)$ and hence the frequency of
oscillations is
\begin{equation}
\omega^{2}=2f_{WC}^{2}(\frac{n}{\Delta n})^{2}\frac{\sigma}{ n m L^{3}_{0}} .
\end{equation}

\section{Drag effect in double electron layers}
\label{drag}

In this section we discuss the more complicated situation in which there are two 2DEG's separated by an effectively impenetrable barrier, but which interact with each other by the Coulomb interaction.    The phase diagram of the two layer system is considerably richer  than that of the single layer, even at mean-field level, due to the possibility of different but strongly coupled patterns of Coulomb frustrated phase separation in each of the layers.  We will discuss a partial solution of this problem in a forthcoming paper \cite{reza2}.

In the present section, we will study the two layer system, because it is the geometry that permits a class of very revealing experiments of the ``drag'' resistance, which allow us to obtain rather direct information about the  character of correlations at intermediate length scales.
The drag resistance, $\rho_{D}=V_{1}/I_{2}$, is the ratio between the longitudinal voltage induced in the upper layer $V_{1}$
by a current in the
lower layer $I_{2}$.  To be concrete,
in this article we will discuss only the case
in which one of the layers is in a bubble liquid phase. Since the mobility of the WC bubbles in the lower layer is finite (even at $T=0$), they move with a velocity proportional to $I_{1}$. Images of the bubbles scatter electrons in the upper layer, thus transfereing momentum to the electron system of that layer.  Generally, at small temperatures this contribution  to the drag
resistance is much bigger than that in the Fermi liquid. Its value depends on the ratio $d/L_{B}$. Specifically, if $d$ is the spacing between layers, the drag effect is large  when the characteristic size,  of the mobile charged objects in the lower layer is greater than or equal to $d$, while relatively little contribution to the drag resistance comes from charge motions which are correlated on length scales small compared to $d$.
Thus, the magnitude of $\rho_D$ is an indirect measure of the characteristic size and magnitude of the charge fluctuations in the system.

\subsection{Drag between two
Fermi liquids.}

As a warmup,  we briefly review some of the  results concerning the drag
between two pure Fermi liquid electron layers.
In this case the electrons  can be described by the
Fermi liquid equations
\begin{equation}
\frac{\partial \tilde{\epsilon}_{\bf k}}{\partial {\bf
k}}\frac{\partial n^{(1,2)}_{{\bf k}}({\bf r})}{\partial {\bf r}} +
\frac{\partial \tilde{\epsilon}}{\partial {\bf k}}\frac{\partial
n^{(1,2)}_{{\bf k}}({\bf r})}{\partial {\bf r}} =I^{1,2}_{st} (n_{{\bf k}})
\end{equation}
\begin{equation}
\label{tildeepsilon}
\tilde{\epsilon}^{(1,2)}_{{\bf k}}=\epsilon_{{\bf k}}+\sum_{{\bf
p}}\xi_{{\bf kk'}}n^{(1,2)}_{{\bf k'}}
\end{equation} where $n^{(1,2)}_{{\bf k}}$ are quasi-particles distribution
functions in the upper and lower layer respectively, $\tilde
\epsilon^{(1,2)}_{{\bf k}}$ are electron energies in the layers,
$\xi_{\bf k,k'}$ are Landau functions, and $I^{1,2}_{st}$ are
scattering integrals describing inter and intra layer electron
scattering.

Let us assume that the electron liquid moves in the lower layer
with a velocity ${\bf U}$. Then the drag resistance is the ratio
between the rate of the change of the momentum in the upper layer
and the current $j=eNU$ in the first layer.

According to Eq.\ref{tildeepsilon} at finite value of ${\bf U}$ the electron
spectrum of quasi-particles in the upper layer $\tilde{\epsilon}_{{\bf k}}$
is anisotropic. Nevetherless at zero temperature the drag
resistance is zero. Indeed, if the quasi-particle distribution
function is an arbitrary function of the total quasi-particle
energy $\tilde{\epsilon}_{{\bf k }}$
\begin{equation}
{\bf J}=\sum_{{\bf k}} \frac{\partial \tilde{\epsilon}_{{\bf
k}}}{\partial {\bf k}}
 n(\tilde{\epsilon}_{{\bf k}})=0
\end{equation}

The drag resistance at finite temperature between two Fermi liquid layers has been analyzed in detail 
in Ref. \cite{MacDonnald}.
So long as the elastic mean-free path, $\ell \gg d$, the drag is given by its value in the absence of disorder ,
\begin{equation}
\label{rhoDFL}
\rho_{D}\sim (k_Fd)^{-4}(k_F/q_{TF})^{2}(T/E_F)^{2},
\end{equation}
where $q_{TF} =2^{-1/2}r_s k_F$ is the inverse of the Thomas Fermi screening length.  Thus, $\rho_D$ increases quadratically with temperature but is small  both because the density of current carrying excitations is small ($\sim T/E_F$) and because the length scale which characterizes the associated fluctuations in the charge density is small compared to the spacing between layers (assuming $k_Fd \gg 1$).  Density fluctuations in the Fermi liquid are enhanced by disorder, with the result that (in addition to a term which is logarithmic in $T$), $\rho_D$, although still small, is enhanced by a factor of $(d/\ell)^2$ in the limit $\ell \ll d$.

The fact that $\rho_D\to 0$ as $T\to 0$
is a property of the Fermi liquid and is not a generic property of all liquids.
To see this, consider the  drag
between two superfluids.  In
this case, the expression for the super-current of the second
layer
\begin{equation}
{\bf J}_{s2}=N_{s2}{\bf U}_{s2}+N_{s12}{\bf U}_{s1}
\label{superdrag}
\end{equation}
has a part which is proportional the the superfluid velocity in
the first layer ${\bf U}_{s1}$ \cite{AndreevBashkin}. Here ${\bf
U}_{sa}$ is the superfluid velocity in the $a^{th}$ layer. The
cross-coefficient $N_{s12}$ in Eq. \ref{superdrag} has a nonzero value at $T=0$.
Thus the system of two super-fluids exhibits non-vanishing drag at $T=0$.

\subsection{Drag between two liquids, one of which is in the
bubble phase.}

 Let us consider the case when the electron densities in the layers
are different. Specifically, let us consider the case in which the ``upper'' layer has a large enough
electron density (small enough $r_s$) that it forms a uniform FL, and thus  plays the same  role as the metallic gate in a MOSFET.
  To begin with, let us consider the case
in which the ``lower'' layer forms a bubble liquid in which  the minority phase (bubbles) is a WC. As
shown in Sec. \ref{bubble}, in this
case there are two types of the current carriers: electrons and
the crystalline bubbles. At zero temperature the electron
Fermi-liquid contribution to the drag resistance is always small and vanishes as $T\to 0$.

Let us consider the bubble phase in the temperature regime in which $\ell_{ee}$ of the upper layer is greater than the spacing between bubbles, $n_B\ell_{ee}^2 \gg 1$.  In this case, transport in the upper layer can be treated using the Boltzman equation.
Each Wigner crystal bubble in the lower layer has an image
of opposite charge in the upper layer. Electrons in the
upper layer scatter on the image potential of the bubble, so the relative motion of the
bubbles in the lower layer leads to a non-zero momentum transfer
between the layers, even at $T=0$.  Specifically, in the absence of disorder pinning, the bubbles drift with the same mean velocity as the rest of the fluid in the lower layer.  Thus,
\begin{equation}
\rho_{D}=\frac{n_{B}}{n}\sigma_{B}
\label{bubbleRD}
\end{equation}
where $\sigma_{B}\sim {\rm max}[d,\rho_B]$ is the cross-section for scattering
the electrons in the upper layer on the image potential from the
bubbles
in the lower layer.  In the presence of disorder, the large, heavy bubbles will  be less mobile than the rest of the fluid in the lower layer, so $\rho_D$ in Eq. \ref{bubbleRD} will be reduced by a factor of the relative mobility of the bubbles.  However, we have shown that bubble localization does not occur, for weak disorder so, although $\rho_D$ may be very small,
even as $T\to 0$ it remains finite.

Other cases, such as the hydrodynamic case in which $\ell_{ee} \ll L_B$, have been discussed in Ref. \onlinecite{SpivakKivelsonDrag}.
The qualitatively important point is that in all cases, $\rho_D \propto n_B$, so $\rho_D$ is directly sensitive to the characteristic
$T$ and
$H_{\|}$-dependences of $n_{B}(T, H)$ which originate from the Pomeranchuk
effect.
The d-dependence of the effect  has also been discussed previously by us  in Ref. \onlinecite{SpivakKivelsonDrag}. At sfficiently small temperatures $T$ the value of $\rho_{D}$ in microemulsions is always larger than its Fermi liquid value and increases as the parameter    $L_B/ d$ increases.

\section{A comparison with experimental results}
\label{comparison}

In this section, we return to the list of experimental signatures of non-FL behavior that we discussed in Sec. \ref{experiment}, and compare them to the behaviors expected on the basis of the  theory  of electronic microemulsion phases. (Recall, the
electron liquid\cite{AbrKravSar,pudalov,krav,sar,kravHpar,pudHpar,okamoto,pudanis,VitSat} in the Si
MOSFET's spans
 a range of $r_s = 5 - 20$ and the hole liquid\cite{Shahar,GaoLanl,Gao2005,Gao2002,GaoHpar05,Pillarisetty,gao01} in p-GaAs heterojunction spans $r_s=10-40$.) We treat the experiments in the same sequence they were presented in Sec. \ref{experiment}.
 In the Section \ref{sarma} we discuss
 problems we see with applying some of the alternative, FL based theories  to the interpretation of these same experiments.

We have established that, in the absence of disorder and for a range of $r_s$ in the neighborhood of a critical
value,
electronic microemulsion phases must
occur.
The most unambiguous experimental evidence of such phases would come from the detection of an accompanying pattern of spontaneous spatial symmetry breaking.
There has been, to date, no direct and unambiguous experimental
 confirmation of these phases at $H_\perp=0$,  although related effects have been observed in the ultraclean 2DEGs in the presence of a large $H_\perp$ \cite{stripeQH}.  However, quenched disorder is always a ``relevant perturbation,'' which makes macroscopic observation of symmetry breaking difficult;  probably, spontaneous symmetry breaking has not been detected in the current generation of large $r_s$ devices  because their mobility is still  insufficiently high.

It is also worth acknowledging that the microscopic aspects of the theory are not, at present, under good quantitative control.  Not only does the characteristic length scale, $L_0$, depend sensitively (exponentially) on unknown microscopic parameters such as the surface tension, but even the range of $r_s$ over which the microemusion phases occur is uncertain.  The numerical calculations that lead to the generally quoted value of $r_s^\star=38$ \cite{cip} do not take into account the fact of the existence of the microemulsion phases and  the enhanced stability of the WC at  finite $T$ due to the spin-entropy ({\it i.e.} the Pomeranchuk effect).
Therefore, even this basic issue must, at present, be treated phenomenologically.

Below we discuss a set of experimental data which indirectly indicate the existence of FL - WC  phase coexistence
 in the strongly correlated  2DEG and 2DHG with large  $r_{s}$.
The theoretical interpretation of these data are based mainly on the Pomeranchuk effect.
As discussed in Section \ref{experfig}, this leads very generally to the qualitative behavior of the resistance as a function of $T$ and $H_{\|}$ shown schematically in 
Figs.  \ref{fig:fig7} and \ref{fig:fig8}.

\begin{figure}
  \centerline{\epsfxsize=10cm \epsfbox{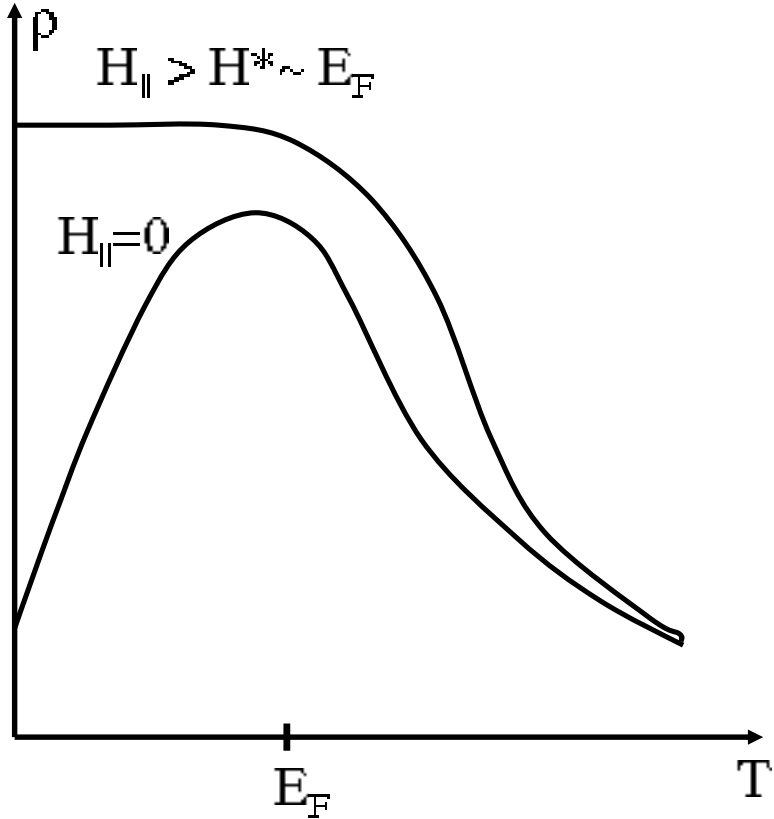}}
  \caption{Schematic representation of the expected temperature dependence of the resistivity of an electronic microemulsion phase in which the FL component is the majority phase.  The
  upper curve corresponds to $H_{\|}>H^{*}$ while the lower curve corresponds to $H_{\|}=0$.
  See discussion in Sec. \ref{experfig}.
 }
   \
  \label{fig:fig7}
\end{figure}

\begin{figure}
  \centerline{\epsfxsize=10cm \epsfbox{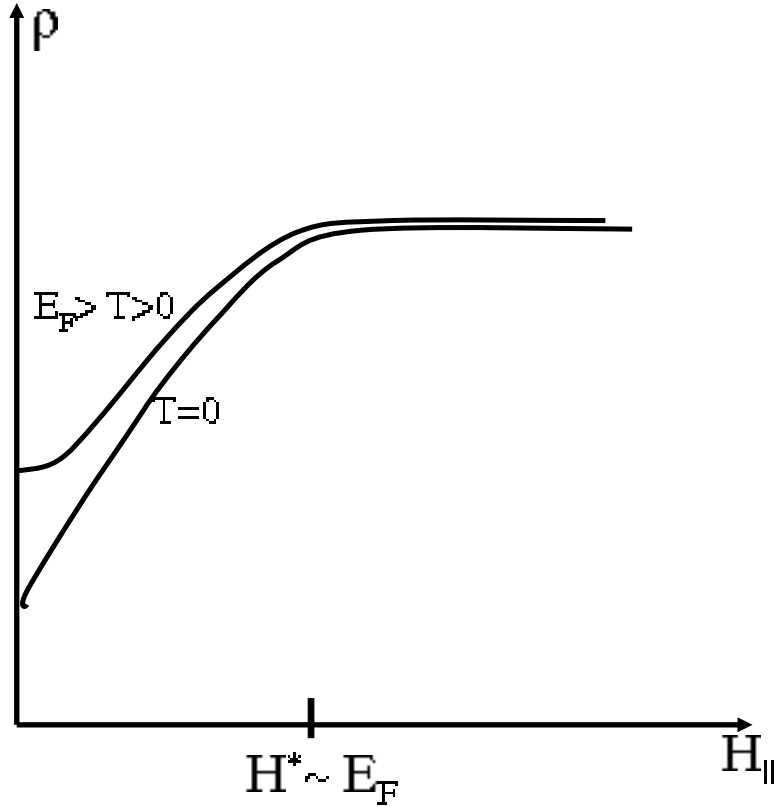}}
  \caption{Schematic representation of the $H_{\|}$- dependence of the resistivity of strongly correlated electron gas under the same circumstances as in Fig. \ref{fig:fig7}.
} \
  \label{fig:fig8}
\end{figure}

\subsection{Experiments on single layers}

 {\bf a.} {\it The existence of an apparent metal-insulator phase transition:}
For zero disorder, there are
several distinct phase transitions between different electronic microemulsion phases which occur as a function of $r_s$.
 However, even weak disorder rounds the symmetry breaking transitions. However, a sharp
 ``percolation'' type transition between conducting and insulating states still occurs   when the WC droplets
overlap and block the electron transport through the FL.
  This, in principle, can qualitatively explain the
 metal-insulator transition observed in \cite{AbrKravSar,pudalov,krav,sar,kravHpar,pudHpar,okamoto,pudanis,VitSat,Shahar,GaoLanl,
Gao2005,Gao2002}.
The Pomeranchuk effect implies that the
critical electron concentration for the
metal-insulator transition is a decreasing function of $H_{\|}$, as is observed in experiment.

{\bf b.} { \it $T$ dependence of $\rho(T,H_{\|}=0)$ at large $r_s$ and $\rho \ll h/e^2$: }
The significant increase of the resistance of metallic samples ( $\rho(T)\ll h/e^{2}$) at relatively
 low temperatures ($T<E_{F}$) with increasing temperature
~\cite{AbrKravSar,pudalov,krav,sar,kravHpar,pudHpar,okamoto,pudanis,Shahar,GaoLanl,Gao2005,Gao2002}
 can be interpreted as being a consequence of the Pomeranchuk effect:
 As the fraction
  of WC increases linearly with increasing $T$, so does $\rho$.
As discussed in Sec. \ref{highT}, at $T\sim E_F$, the  WC melts and a crossover occurs to
 a semiquantum regime in which the viscosity Eq. \ref{andreev}. and consequently,
  the resistance  fall in proportion to $1/T$
   or faster (Eq. \ref{classicaleta}).
     Thus, theoretical expectations are consistent with the fact that $\rho(T)$ reaches a maximum
     and then decreases when $T>E_{F}$.

{\bf c}.  {\it $H_{\|}$ dependence of $\rho(T\approx 0,H_{\|})$ at large $r_s$ and $\rho \ll h/e^2$:}
The Pomeranchuk effect   implies that the low temperature resistance of the 2DEG increases linearly with $H_{\|}$ for  $H_{\|}< E_{F}$
and saturates when $H_{\|}>E_{F}$, where complete spin polarization is achieved.
All this is in good qualitative agreement with the experimental observations in
Refs. \cite{AbrKravSar,pudalov,krav,sar,kravHpar,pudHpar,okamoto,pudanis,VitSat,
GaoHpar05}.
For small $H_{\|}$ there should be a scaling relation between the $T$ and $H_{\|}$ dependences implied by Eq. \ref{pomWC} which has not, to the best  of our knowledge, been critically tested in experiment.

{\bf d.}  {\it The quenching of the $T$ dependence of $\rho(T,H_{\|})$ for $H_{\|} > H_{\|}^\star$
at large $r_s$ and $\rho \ll h/e^2$:}
Close enough to the FL side of the microemulsion phases, the FL portion of the system can percolate even when the fraction of WC has been enhanced by the application of $H_{\|} > H_{\|}^\star$ .  Thus,
far enough from the ``metal-insulator transition,''  we expect a range of densities in which the resistance increases strongly as a function of $H_{\|}$, but still
has a metallic value at high field: $\hbar/e^{2}\gg \rho(H_{\|}^{\star})\gg \rho(H_{\|}=0)$.
However, once all the spins are polarized (for $H_{\|}> H_{\|}^\star$), the Pomeranchuk effect is quenched, and consequently
so is its contribution to the
$T$-dependence of $\rho(T)$.
Thus, in our opinion, the most direct confirmation of the central role played by the Pomeranchuk effect
comes from a comparison of the temperature slopes of the resistance with and without the magnetic field.
 The $T$ dependence of $\rho$ is found to be
1-2 orders of magnitude weaker  for $H_{\|} > H^\star$ than for $H_{\|}=0$
 and can even change sign.~\cite{VitSat,GaoHpar05}
Such a dramatic effect strongly suggests that $H_{\|}$ completely quenches the dominant contribution to the $T$
 dependence.

{\bf e. }  { \it ``Metallic'' $T$ dependence of $\rho(T,H_{\|}=0)$ at large $r_s$ and $\rho > h/e^2$:}
The conventional picture of the disorder driven metal-insulator transition
implies that samples with large resistivity, $\rho\gg h/e^{2}$, at low temperatures
should exhibit hopping conductivity, so  $\rho$ is expected to increase strongly as the temperature
decreases.
However, there are cases (See for example Fig.1 in
 Ref. \cite{gao01}) in which highly resistive 2DEGs ($\rho \sim (2-3) h/e^2$) with large $r_s$ violate this expectation, in that $\rho$ is an increasing function of {\it increasing} $T$.  If what is changing with $T$ is not a scattering rate, but the fraction of the system that is highly conducting, then there is no maximum value of the resistivity for which such metallic $T$ dependence is possible.  This
observation may have much wider implications for the so-called\cite{emeryprl} ``bad metal'' behavior observed in many highly correlated materials, for which related\cite{hebbard,ando}  intuitive explanations have been proposed.

{\bf f.}  {\it Oscillations of  $\rho(H_{\bot})$ at $T>E_{F}$ }.
The existence of the oscillations of the resistance $\rho(H_{\bot})$
 as a function of $H_{\bot}$ at $\Omega_{p}\gg T\gg E_{F}$ \cite{GaoLanl} is an experimental
  fact which drastically contradicts FL theory. At these temperatures, the WC has
   melted and the system is in the semiquantum regime.
Although we do not have a theory explaining this phenomenon, we suspect that it reflects
directly the strongly correlated nature of the electronic system:  Since the kinetic
 energy in the electronic system is still much smaller than the potential, on short
  spacial scales the liquid may have correlations which are similar
to those of a crystal. Thus, in the interval of magnetic fields  where  the magnetic
 length is comparable to
the interelectron distance, short range quaisiperiodic order may account for a few
oscillations of $\rho(H_{\bot})$.

\subsection{Experiments on double-layers}

We now turn to the anomalies in the measured
 drag resistance $\rho_{D}$ \cite{Pillarisetty} in p-GaAs double layers
 with small resistances $\rho\sim (1/10- 1/20)h/e^{2}$ and large $r_{s}$.

{\bf a.} {\it The magnitude of $\rho_D$}:  The fact that $\rho_{D}$ is 1-2
order of magnitude larger than expected on the basis of Fermi
liquid theory is in  qualitative agreement with our
theory.  Specifically, where the FL is the majority phase in the lower  layer, the fact that the expected size of the WC domains ( Eq. \ref{L0}) is $L_0\ge d$, where $d$ is the spacing between layers, means that none of the familiar factors responsible for the small size of $\rho_D$ in a FL (Eq. \ref{rhoDFL}) apply.  Put another way, the large magnitude of $\rho_D$ can be viewed as direct confirmation of the existence of dynamical charge inhomogeneities of substantial magnitude and on length scales greater than or equal to $d$.

{\bf b}, {\bf c}, and {\bf d}. {\it $T$ and $H_{\|}$ dependence of $\rho_D(T,H_{\|})$, and its similarity to $\rho(T,H_{\|})$}:  In addition to its large magnitude, the most striking feature of the measured $\rho_D$ in large $r_s$ double layer devices is that the  $T$ and especially the $H_{\|}$ dependencies
looks
qualitatively similar to
the resistances of the individual
   layers $\rho(T,H_{\|})$.
Most notable is the fact that, again, a strong magnetic field, $H_{\|}> H^\star$
 substantially suppresses the temperature dependence
of $\rho_{D}(T)$ \cite{Pillarisetty},
in qualitative agreement with the assumption that the dominant $T$ dependence
 is a consequence of the Pomeranchuk effect.
More generally, the qualitatively similarity between $\rho$ and $\rho_D$
speaks in favor of the assumption
 that properties of
both $\rho(T,H_{\|})$  and $\rho_{D}(T, H_{\|})$ are connected with
 phase separation.

\subsection{Alternative theories
 of experiments in large $r_s$ devices
  }
\label{sarma}

 Attempts have also been made (Refs.~\cite{dolgopolov1,SarmaHwang,Aleiner}) to explain the observed behaviors by extrapolating known perturbative expressions which are valid in the limit of small $r_s$ and $\rho \ll h/e^{2}$, to the regime of large $r_s$.
In this section, we point out some of the difficulties with these FL based explanations.
  Before getting to specifics, however, it is worth noting from the beginning that there is a general
   problem with applying  these theories to the experiments discussed above:
Perturbative corrections are, by their nature, small compared with the resistance itself, while the observed
 variations of $\rho$ with $H_{\|}$ and $T$ are large - up to a factor of 10.  Moreover, in all cases,
to explain the experimental data, the FL results are being extrapolated far beyond their
natural range of validity -- specifically, $r_s$ is much greater than 1.

In the weakly interacting limit, $r_{s}\ll 1$, and at $k_{F}\ell\gg 1$, calculations of the
resistance of the 2DEG  can be
carried out in a controlled fashion.
To lowest order in powers of  $(k_{F}\ell)^{-1}\ll 1$, the transport
properties of the system can be described by the Boltzmann kinetic
equation and the Drude formula for the resistance, which at low $T$
is determined by the impurity scattering. The $T$ and $H_{\|}$
dependences of the resistance $\rho(T,H_{\|})$ are determined by
the corresponding dependencies of the density of states at the Fermi
level, the Fermi velocity, and the scattering cross-section, and
typically $\rho(T, H_{\|})$ decreases  weakly with $T$ and $H_{\|}$ because
the electron-impurity cross-section is a decreasing function of the
electron energy.  Even the sign  of the
effect is opposite to what is observed in large $r_s$ devices.

If $\rho\ll h/e^{2}$, and at moderately low temperatures, 
 both weak localization ~\cite{Abrahams,khmelnitski} and
 the electron interaction interference corrections ~\cite{Altshuler} to the classical
 Drude resistance are parametrically small.
 The physics of the interactional corrections 
 is the following:
Impurities in a metal create  Friedel oscillations of the electron
density. Due to the electron-electron interactions, the
quasiparticles in the
 metal are scattered
not only from the impurity but also from the modulations of the
 electron density.
The interference between multiple electron scattering on impurities
and on the Friedel oscillations of the electron density give rise to
the non-analytical in $T$ and $H_{\|}$ corrections to the Drude
resistance.  At high enough temperatures when $L_{T}=v_{F}/T \gg l$ the
former effect is reduced to an interference between  two scattering
amplitudes: The electron  scatters from a single short range impurity
from the  Friedel oscillations created by the same
impurity. The
interference corrections to the Drude formula in this so called
``ballistic regime''  were considered in
~\cite{stern,Sternsarma,dolgopolov1,Aleiner}. 
 The
result is that in an intermediate interval of temperatures
 $E_{F}\gg T\gg 1/\tau_{i}$, the $T$- dependent correction
to the resistivity $(\rho(T)-\rho(0))\sim  C T$  is linear
 in $T$ and $C\sim F_{0}(1+15F_{0})$ ~\cite{Aleiner}. Here $F_{0}$ is
 a Fermi liquid parameter which,  for $r_{s}\ll 1$, is negative with a magnitude proportional to $r_{s}$.
 Thus, at small $r_{s}$, when $F_{0}$ is small,   $C < 0$ and $\rho(T)$ decreases as $T$ increases.
 However, at large enough $r_{s}$, when $F_{0}<-1/15$, the coefficient $C$ changes sign,
 and $\rho(T)$ becomes an increasing function of $T$, which is in agreement with the experiments.
 In principle, at $r_{s}\sim 1$ and $H_{\|},T\sim E_{F}$ the corrections could be
 of order one.
It is hard, however, to imagine that the  Friedel oscillations of the electron density induced by an
 impurity can make a much larger contribution to the scattering cross-section than the scattering from the
  impurity itself. Therefore, we think that this process (even when extrapolated to large $r_s$)
   cannot explain changes in the resistivity by more than a factor of two, as are observed experimentally.

Another aspect of the problem is that the calculations of Ref.~\cite{Aleiner} are valid only for
 $T$ and $ H_{\|}\ll E_{F}$, where one can neglect the energy  dependence of the electron
  scattering cross-section. In addition, for $r_{s}\ll 1$, where the theory is well controlled,
    at $ H_{\|}>E_{F}$, or at $T\sim E_{F}$, the main contribution to the magnetoresistance
    $\rho(H)-\rho(0)$   comes not from Friedel oscillations considered in ~\cite{Aleiner}, but rather
     from the $H_{\|}$ dependence of the Drude part of the resistance.
    At last the theory Ref. ~\cite{stern,Sternsarma,dolgopolov1,Aleiner} can not explain 
     why the $T$-dependence of the resistance $\rho(T)$ is completely quenched at $H_{\|}>H^{*}$.
Therefore we think that, though the results of the perturbation theory ~\cite{stern,Sternsarma,dolgopolov1,Aleiner} may be relevant to experiments
performed on samples with relatively small $r_{s}$ and at relatively small values of $H_{\|}$ and $T$, they can not explain results of experiments at $r_{s}\gg 1$ where almost all effects are larger than order one.

An attempt at explaining results
on large $r_s$ samples was undertaken in Ref. \onlinecite{SarmaHwang}  using the classical formulas
 for the resistivity and extending them to the regime $r_{s}\gg 1$ and $k_F\ell \sim 1$.
   There are several aspects of this extrapolation that are troubling.  Firstly, as a function
    of $r_s$, both $d \rho/ dT$ and $d \rho /d H_{\|}$ change sign, so the extrapolation cannot
    be said to be featureless.  Moreover, the origin of the sign change can be traced to the
    fact that at large $r_s$, the Thomas Fermi screening length is parametrically smaller than
    the spacing between electrons, $k_{TF} \sim
{r_s} k_F\gg k_F$.  Screening lengths less than the distance between electrons are clearly unphysical.
Moreover, the calculation treats the impurity scattering in Born approximation, which is
also unjustified in this case.
On a phenomenological level,
the extreme degree to which the magnitude of $d\rho/dT$ is suppressed by  large $H_{\|}$
 in both p-GaAs and Si MOSFETs requires an unreasonable amount of fine tuning of parameters.

 \section{conclusion}
\label{conclusion}

In this section we discuss several important unsolved problems
related to  Coulomb frustrated phase separation.

\subsection{Unsolved problems concerning the equilibrium phase diagram}

{\bf a}
As discussed in Sec. \ref{quantumbubble}, quantum fluctuations necessarily melt a bubble crystal when
 the spacing between bubbles is sufficiently large. Thus, the mean-field Lifshitz transition between
  the uniform fluid and the bubble crystal is destroyed by quantum fluctuations; the
bubble crystal phase is only stable when the lattice constant
is smaller than a finite critical value.  However, any transition
from a liquid phase to a crystal with finite period is expected to
be first order,
 in conflict with the theorem forbidding the existence of first order transitions.
This is a conceptual problem that we have failed to resolve.

 A related unresolved problem is the quantum ($T=0$) properties of the melted WC bubble phase.

{\bf b}. The properties of the interface between the FL and the WC are largely unknown.
  The question of whether the interface is quantum smooth or quantum rough is open.
 In addition, the characteristic periods of the two phases 
 are different - $2\pi/2k_F$ for the FL and the lattice spacing for the WC, opening the possibility of interesting effects of incommensurability.   The interfaces may even have phase transitions of their own.

\subsection{Significance of quenched disorder}

The strongly correlated liquid in the presence
of an external scattering potential is a largely unsolved problem. There is a theorem forbidding  first order phase transitions in
2D with arbitrarily small disorder \cite{Imry}.
In the case of weak disorder, though, the characteristic length
at which the first order phase transition is destroyed can be
exponentially large.
According to \cite{Imry}, whether the first order phase
transition is converted to a second order one, or to a cross-over depends on
whether the two phases have different symmetries.
However, long range
crystalline order is impossible in the presence of quenched
disorder
in dimensionalities  less than four \cite{Larkin};  the result is probably,
a glassy state.

One can see the glassy aspects of the
problem even on the mean field level 
if one assumes that the coefficients in Eq. \ref{Fmicro1} are smooth functions
of the coordinates.  So long as $L_0\gg\xi$, a coarse-grained
description of the thermodynamic state of the system can be obtained by
associating an ``up'' psuedo-spin with each local position which is in the
WC state, and a ``down'' with each FL position.  The problem of finding the
mean-field ground-state then corresponds to the same problem in an effective
random-field Ising model, albeit one with power-law interactions.  While
there is little that is actually known about the random field Ising model with longer
range interactions, it is certainly likely that, in common with  the
short-range model, it forms a glass.

{\it Implications for localization theory in 2D conductors:}
An important open issue concerns the implications of Coulomb frustrated phase separation for
 the theory of localization in the presence of weak disorder.
It is believed that at $T=0$, all electron wave functions in
disordered systems are localized. This statement is based on the
logarithmic divergence of the interference corrections to the Drude
conductivity\cite{Abrahams,khmelnitski,Altshuler} and its interpretation via  the
renormalization group \cite{Abrahams}. In turn, these
results are intimately connected to the fact that, in the presence
of an elastic scattering potential, the number of quasiparticles
with a given energy is  conserved.

In the case when there is a finite concentration of WC droplets in the FL, the status of
 this statement is unclear. As discussed in Sec.
\ref{quantumbubble}, due to quantum fluctuations, the bubbles can  quantum
melt and recrystalize. Thus the number of electrons in the Fermi
liquid is not conserved because  they can join the WC fraction of
the system.
Consequences of this fact for the 2D localization theory remain to
be investigated.

\subsection{Mesocopic properties}

An important area of research involves transport through metallic ``point contacts.''
Up until now, however, they have been studied only in three regimes: the regime where electron
interactions are not important \cite{Wees}, the
Coulomb blockade regime,   and
the Kondo regime \cite{RikhGlazmanGoldhaberGordon} in which
there is a single localized electronic state through which tunneling between the two metallic reservoirs occurs.
 However, there is another interesting regime when there is a relatively large ``depletion region''
 between two metallic reservoirs with high electron densities.  In the depletion region, there is a
 strongly correlated electron liquid with low electron density whose density can be varied by
 changing a gate voltage.
In this case as a function of gate voltage, aspects of the
electronic micro-emulsion phases can be directly probed on a
mesoscopic scale.

 In this context we would like to mention several problems:

 a) The Pomeranchuk effect can produce significant $T$ and $H_{\|}$
 dependencies to the resistivity through the depletion region. Suppose that at $T=0$ the system is in the FL
 state, but with the gate tuned so that the WC is close in energy.
Then, as the temperature or the magnetic field increase, the resistivity can exhibit a sharp crossover from very conducting to very insulating values.
  This can mimic some of the behavior traditionally associated with the Kondo effect.

b) An example of a mesoscopic geometry which is well suited to
exposing electronic micro-emulsions is one
 in which the depletion region is produced by
a smooth external potential which depends only on one coordinate.
This ``gutter-like''  potential is shown qualitatively
in Fig. \ref{fig:fig10}.
Specifically, it is interesting to consider the nature of the transition from the insulating state
 when the entire gutter is filled with low density WC, to a conducting state in which the highest
 electron density portion of the system, which lies along the line of minimum potential at the
 center of the gutter, is a liquid.
At $T=0$, in the limit that the gutter is long, this is a one dimensional quantum phase transition.

\begin{figure}
  \centerline{\epsfxsize=10cm \epsfbox{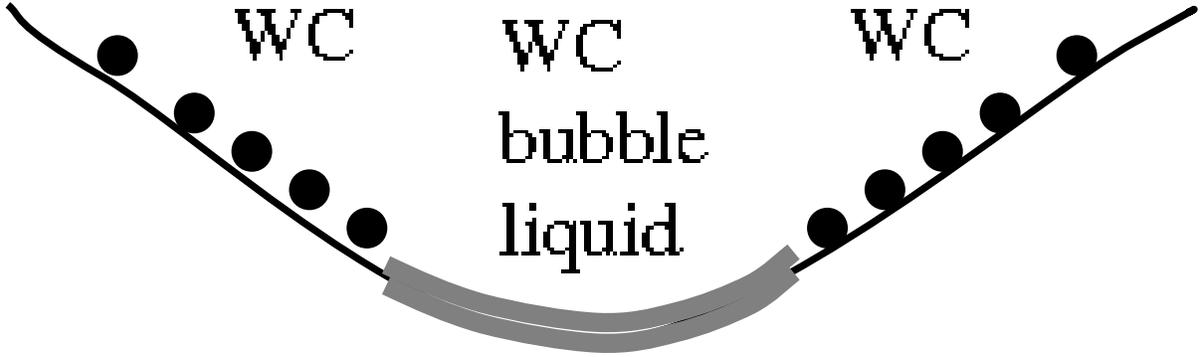}}
  \caption{A schematic picture of a smooth gutter like potential in a split-gate device.
  The dots indicate electrons in the WC. The gray area corresponds to
  a melted FL bubble phase.} \
  \label{fig:fig10}
\end{figure}

At mean-field level, the first transition that occurs with increasing electron
density is to a phase with far separated FL bubbles;  because the bubbles have relatively higher
 electron density than the surrounding WC, they are bound to the center of the gutter forming a 1D bubble crystal.  However,  quantum fluctuations of the bubble's positions will always cause the  bubble crystal
   to melt.
 The resulting bubble liquid phase is thus a conducting (fluid) phase.

The bubble motion also produces changes in the spin-state of the surrounding WC.  As discussed in Sec. \ref{quantumbubble}, 
the FL
bubbles are point like
 quantum-mechanical
objects whose zero-point kinetic energy is lowest when the surrounding spin's are ferromagnetically polarized.  Thus, as discussed in the neighborhood of Eq. 47, if the ratio of the bubble kinetic energy to the exchange energy in the WC is large enough, zero temperature kinetically stabilized ferromagnetism can onset with the onset of the bubble liquid phase.
In this context, we note that there are a set of experiments
Ref.~\cite{Peper1,Peper2,Peper3,Peper4} in various mesoscopic devices,
 with a geometry similar to that of the gutter-like potential we have just
discussed, in which a set
 of anomalies referred to as the ``0.7 effect'' have been observed.
  In Ref.~\cite{Peper1}, this anomaly was interpreted as evidence 
of the existence of ferromagnetism in 1d channels at
 large $r_s$.
 
c) Another interesting but still incompletely understood issue
involves
 a particular type of quantum oscillations of the WC droplets of
the sort illustrated in Fig. \ref{fig:fig9}.  Here the black dots
represent electrons in a crystal droplet which is bound at a
preferred position by the disorder potential.  Consider a process in
which
 the bubble  sloshes a bit to the right, then an appropriate sized sliver of
 the  displaced crystal  quantum melts, and an equivalent sliver of
liquid coherently recrystallizes on the left.  The result is that
the bubble has returned
 to its original position.  The value of the action corresponding to
such a process is finite.
An important feature of the process, however, is that there is net
charge motion that accompanies it.

\begin{figure}
  \centerline{\epsfxsize=10cm \epsfbox{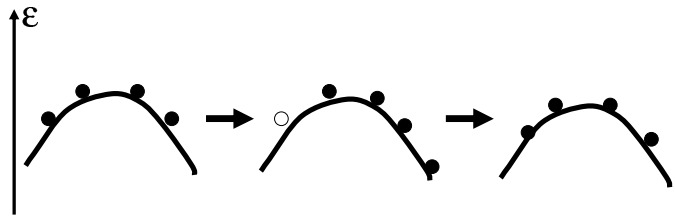}}
  \caption{
  Schematic representation of the periodic oscillations of a WC droplet bound to a potential ``mountain-top'' which results in quantum pumping action, detectable in drag.} \
  \label{fig:fig9}
\end{figure}

 In equilibrium, oscillations of the droplet are equally probable both
to the right and to the left.  However, the problem can be biased by
application of a
 voltage.  Then, this effect can be detected in
 drag:  as the droplet in the lower plane executes this quantum motion, the image
 charge in the upper layer also acts as a pump of electrons from
the left to the right.
 This gives additional support to the idea of a non-zero drag
 resistance in the limit $T\to 0$.

{\it Note added:}  After we completed this article, we became aware of a paper, \onlinecite{manousakis}, in which the related problem of Coulomb frustrated phase separation in a bosonic system was studied by quantum Monte-Carlo methods.  Here stripe and bubble crystal phases are clearly seen, although various subtleties, such as  the hierarchical microemulsions at that we would expect between the stripe and bubble phases, have not yet been explored.

This work was supported in part by the National Science
Foundation under Contracts No. DMR-01-10329 (SAK) and DMR-0228104
(BS).  We thank L. Levitov, R. Pillarisetty, B. Shklovskii, D.
Tsui,S.Das Sarma, S. Maslov, V. Pudalov, S. Kravchenko, M. Sarachik, S. Vitakalov, G. Boebinger, X.P.A. Gao, and D. Shahar for useful discussions.

\appendix

\section{Coulomb frustrated phase separation}
\label{thermoappend}

 In this appendix we present a more extended analysis of the results 
obtained in Refs. \cite{Reza,SpivakPS,SpivakKivelsonPS}, give qualitative explanation of existence of the phase separation on the mesoscopic spacial scale (microemulsions), and derive  Eqs. \ref{Fmicro} and \ref{Fmicro1}.

\subsubsection{ Qualitative Features}

A putative first order transition from the WC to the FL would occur when the free energy densities of the two uniform phases, $F_{WC}(n)$ and $F_{FL}(n)$, cross, as in Fig.  \ref{fig:fig2}.    In the case in which $V(r)$ is long-range, {\it i.e.} it falls as $1/r^2$ or slower, $F(n)$ must include the interaction with a uniform positively charged background
which is assumed to be static and uniform.
As a first step, we show that for $n$ near $n_c$, neither a uniform phase, nor a macroscopically phase-separated combination of them, minimizes the free energy - rather some form of mesocscale phase separation always has lower free energy.

To see the point most simply, consider the system with average density $n=n_c$.  We divide the system up into alternating stripes of WC and FL of width $L \gg \xi$, where $\xi$ is the width of the interface between a region of WC and FL.
( We consider stripes to be explicit, but other patterns, such as an array of bubbles of radius $L$, have similarly low free energies.)
We also rearrange the density, so the WC stripe has density $(n_c - \Delta n)$ and the FL stripe has density $(n_c+\Delta n)$.  The free energy density of this state can be expressed as a sum of different contributions
\be
F= F_{un}+F_{ps}+ F_{\sigma}+F_{lr}.
\label{Femulsion}
\ee
The free energy density of the  uniform density phase is
$F_{un}=F_{WC}(n_c)=F_{FL}(n_c)$.
The next term is what would have been the gain from phase separation, were there no long-range interactions
\ba
\label{Fps}
F_{ps}=&&(1/2)[F_{FL}(n_{c}+\Delta n)-F_{WC}(n_{c}-\Delta n)] \\
&&= -\zeta_{1}\Delta n + \zeta_{2}(\Delta n)^2 + \ldots
\nonumber
\ea
Here $\zeta_{1}$ and $\zeta_2$ are the first terms in an expansion in which    $\ldots$ signifies higher order terms in powers of $\Delta n$.
The third term in Eq. 1  is the short-distance contribution to the surface energy,
\be
F_{\sigma}=\sigma/L
\ee
 where $\sigma$ is the surface tension.
The final term is the  the long-range interaction
in this particular trial state:
\be
 F_{lr}=(1/2\Omega) \int d {\bf  r} \ d {\bf  r}^{\prime} V({\bf  r}- {\bf r}^{\prime}) \delta n({\bf  r})\delta n({\bf r}^{\prime}).
\label{FCoul}
\ee
where the difference between the local density and the average density $\delta n({\bf  r})=\pm \Delta n$ depending on whether one is in a WC or FL portion of the system, $\Omega$ is the total area of the system, and $\int d  {\bf  r} \ \delta n({\bf  r}) = 0$.
Though the cases $x=1$ (Coulmb case) and $x=3$ (the dipolar case) are the most interesting, physically, we start with the case $1<x<3$ because the result is more easily established than in the limiting cases.

The integral in Eq. \ref{FCoul} has a different character depending on whether $x$ is greater or less than $x=2$.

a) {\it The case $1<x<2$.} In this case  the integral is evaluated as:
\be
F_{lr} = A Q (\Delta n)^2L^{2-x} \ \ {\rm for } \ 1 < x < 2,
\ee
where $A$ is a positive number of order 1.  Notice that  the energy density diverges as $L\to \infty$, reflecting the fact that macroscopic phase separation is impossible with long-range interactions.
Consequently, if  $\Delta n$ is small (so that the higher order terms in $F_{ps}$ can be ignored) and $L$ large compared to $\xi$,
\be
F = F_{un} -\zeta_1 \Delta n +\sigma_0/L + AQ(\Delta n)^2L^{2-x}.
\ee
This expression can now be minimized with respect to $\Delta n$ with the result
\be
\Delta n=(\zeta_{1}/2AQ)L^{-(2-x)}; \ \   F=F_{un}-L^{-1}\left[ (\zeta_1^2/4AQ)L^{x-1} - \sigma\right] .
 \label{Delta1}
 \ee
Notice that the long-range interaction produces something resembling a scale-dependent surface tension, which is negative for large enough $L$.  Minimizing this expression with respect to $L$, we find the characteristic scale for nano phase-separation,
\be
  L=L_0 \equiv[4AQ\sigma/(2-x)\zeta_{1}^2]^{1/(x-1)} .
  \label{L1}
  \ee
This expression is valid if $\sigma$, and hence $L_0$ is sufficiently large, so that $\Delta n$ is small;  if $\sigma$ is not large, the result depends on the higher order terms in $F_{ps}$, although this typically will not change the result qualitatively.

{\it b) The  case, $2 < x \le 3$:}
 In this case both the long and short-distance contributions to the integral are singular, with the consequence that
\be
F_{lr}= Q(\Delta n)^2\left[A^{\prime} \xi^{-(x-2)} - AL^{-(x-2)} +\ldots\right]    \ \ \ {\rm for} \ x > 2
\label{MR}
\ee
The first term can be thought of as a positive contribution to $F_{ps}$, $\zeta_2\to\zeta_2 + AQ\xi^{-(x-2)}$,
while the latter can be thought of as a negative, but again scale dependent contribution to the surface
 tension, $\sigma \to \sigma-AQ(\Delta n)^2L^{3-x}$.

In this case, macroscopic phase separation is possible, and the density difference, $\Delta n^{\star}$   between
 the two coexisting phases is determined by minimizing
  $\tilde F_{ps}(\Delta n) \equiv F_{ps}(\Delta n) + A^\prime Q(\Delta n)^2\xi^{2-x} $
   with respect to $\Delta n$ -- {\it i.e.} by the Maxwell construction.
  The free energy of the macroscopically phase separated state that results is then
   $F_{un}+\tilde F_{ps}(\Delta n^{\star}) < F_{un}$.  However,  taking $L$ to be large but non-infinite,
    results in a state with free energy density:
\be
F=F_{un} +\tilde F_{ps}(\Delta n^{\star})- L^{-1}\left [ AQ(\Delta n^{\star})^2L^{3-x} - \sigma\right ]
\label{F2}
\ee
which is manifestly lower than the free energy of the macroscopically phase separated phase for any
$L > [\sigma /AQ(\Delta n^{\star})^2]^{1/(3-x)}$.  The optimal domain size is
\be
L=L_0\equiv [\sigma /(2-x)AQ(\Delta n^{\star})^2]^{1/(3-x)}.
\label{L2}
\ee

The cases of $x=1$ and $x=3$ are discussed in detail in
\cite{Reza,SpivakPS,SpivakKivelsonPS}. They require a more subtle
analysis in the sense that it is necessary to consider a form of
phase separation in which the density varies within each domain.
For instance, for $x=1$, in the low density (WC) domain, the density
varies as $n_c-\Delta n/r$ as a function of the distance, $r$, from
the interface with the high density phase.  When this is done, the
result is to replace $L^{x-1}$ in Eq. \ref{Delta1}, $L^{3-x}$ in Eq.
\ref{F2}  by $\log[L/\xi]$ and  the expressions in Eq. \ref{L1} by
\be 
L=L_0 \equiv  \xi \exp[4AQ\sigma/\zeta_1^2+1]. 
\label{L02} 
\ee
Here the short distance cut off, $\xi$ is of order $d$ in the
dipolar case and of order ineterelectron distance  in the Coulomb
case.  (This result is equivalent to Eq. 4.)

It is worth noting that, in the Coulomb case,  the micro-emulsion phase has lower free energy
 than the uniform phase, while for the dipolar case, the competition is between the
  macroscopically phase separated phase and the microemulsion.  However, this is the
   only difference -- the same logarithms enter the free energy expression for the
    microemulsion phase in both cases, and hence the character of the microemulsion is,
     surprisingly, similar for Coulomb and dipolar interactions.

 \subsubsection{Derivation of Eqs. \ref{Fmicro} and \ref{Fmicro1}}
 
So as to be able to obtain explicit results, we consider an effective free energy density describing a first order phase transition
 between two phases, a high density phase, in which a suitably defined order parameter $\phi({\bf r})=1$,
  and a low density phase in which $\phi({\bf r})=-1$: 
 \be
 F=\Gamma[\phi ]-\zeta (n({\bf r})-n_{c})\phi({\bf r})+\frac{1}{2}(n({\bf r})-\bar n)
 \int d {\bf r'} V({\bf r}-{\bf r'}) (n({\bf r'})-n).
 \label{FmicroDer}
 \ee
Here
$\Gamma[\phi ]$ is a  local functional of $\phi$ with two degenerate minima corresponding to $\phi({\bf r})=\pm 1$.  It also has a saddle point  corresponding to a domain wall between the two phases of width $\xi$ and energy per unit length $\sigma > 0$.    The final terms determine the long wave-length profile of the density,  $n(\bf r)$, where $\bar n$ is a fixed uniform background density, and $n_c$ is the critical density.  (Where the optimal density of the two phases is very different, higher than quadratic terms in powers of $(n({\bf r}) - \bar n)$ may be quantitatively  but not qualitatively important.)

Minimizing  Eq. \ref{FmicroDer} with respect to $n({\bf r})$ we get
 \be
 n({\bf r})-\bar n= \zeta V^{-1}\phi({\bf r}),
 \label{nprofile}
 \ee
 where $V^{-1}$ is the operator inverse of $V$, which
 can be defined  by Fourier transform form as
 $(n({\bf r})-\bar n)_{{\bf k}}=\zeta (1/V_{{\bf k}}) \phi_{k}$.   Substituting this expression back into Eq. \ref{FmicroDer}
 we get an expression for total energy of a microemulsion phase 
 \be
 F_{ME} = \int d{\bf r} \left \{\tilde \Gamma[\phi({\bf r})]-\zeta (\bar n-n_{c})\phi({\bf r})
 -(1/2)\int d {\bf r'} \partial_{{\bf r}} \phi({\bf r})  \tilde {V}({\bf r}-{\bf r'}) \partial_{{\bf r'}}\phi({\bf r'})  \right \} .
 \label{FmicroDer1}
 \ee
 For $1 \le x\le 2$, since $V_{\bf k}\to\infty$ as $|{\bf k}| \to 0$,  the dominant effect of the long-range interactions is expressible as a non-local stiffness, so  $\tilde \Gamma = \Gamma$ and $\tilde{V}(r)\sim\zeta^2 Q^{-1}/|r|^{2-x}$.
 For $2 < x \le 3$, the dominant contribution is local, resulting in a renormalization of $\Gamma \to \tilde \Gamma = \Gamma - (\zeta^2/2V_{\bf 0}) \phi^2$, and it is the subleading term which  produces a non-local stiffness, $\tilde{V}(r)\sim \zeta^2Q^{-1}/|r|^{x-2}$.
 The amusing result is that the  form of the free energy depends only on $|x-2|$, and in particular that $\tilde{V}(r) \sim 1/r$ both for the Coulomb case, $x=1$, and the dipolar case, $x=3$.

  In the limit when the widths of the interphase is smaller than the size of domains one can neglect the gradient of $\phi({\bf r})$ inside the phases and we arrive to Eqs. \ref{Fmicro} and \ref{Fmicro1}.

\section{Derivation of Eq. \ref{tij}}
\label{delocalization}

 Below we present a heuristic derivation of Eq. \ref{tij} but the same result can be derived more formally following
 the methods of Levitov and Shytov\cite{LevitovShytov}.

To understand the origin of the non-local tunneling represented by Eq. \ref{tij}, we begin by considering a process in which a bubble of WC at site $j$ quantum melts.  (We will find that this process has a logarithmically divergent action -- were the action finite, it would lead to a term proportional to $[a_i^\dagger+a_i]$ in Eq. \ref{tij}.)  We imagine that there are two nearly degenerate states of the system, one in which there is a bubble at site $j$, and one with a slightly different energy, $-\epsilon_j$, in which site $j$ is covered with liquid.  Since in equilibrium, the liquid has a slightly higher density than the WC, the transition between the two states must involve the motion of an amount of charge $Q=e\Delta n L_B^2$ in from ``infinity,'' {\it i.e.} from a distance $L_\infty$ which is the system size.  The classical action for this process is thus the sum of three contributions:
\begin{equation}
S=S_{melt}+S_{hydro}+S_{\epsilon} .
\end{equation}
where, as in the usual theory of quantum nucleation, 
\be
S_{melt} \sim \sigma L_B^2/v_F
\ee
is the action involved in melting the WC bubble locally,  $S_{hydro}$ is the action associated with the motion of a charge $Q$ in from infinity, and (with 
$\epsilon\equiv -\epsilon_j$)
\be
S_\epsilon =-\int d\tau \epsilon
\label{Seps}
\ee
records the cost of leaving the system in a higher energy metastable state.  Since the charge redistribution takes place entirely in the FL and dominantly at long distances, $S_{hydro}$ depends only on the hydrodynamics of the FL.  In Eq. \ref{Seps}, the integral runs over the ``imaginary'' time interval which characterizes the event.  As we shall see, both $S_{hydro}$ and $S_\epsilon$ diverge as $L_\infty \to \infty$.  However, when applied to the process of melting and recrystallization at a site a distance $R$ away, the same analysis leads to a finite result, with $L_\infty$ replaced by $R$.

  A lower bound estimate of $S_{hydro} $ can be made by integrating over time the instantaneous excess energy associated with the charge excess,
\begin{equation}
S_{hydro}\approx \int _0^\infty d\tau E(\tau).
\end{equation}
In the short-range case (including the dipolar case),
\begin{eqnarray}
E(\tau)=(1/2)\kappa \int d^2r [\delta n(r,t)]^2 \sim \kappa [\Delta n]^2[L_B/L(t)]^2
\nonumber
\end{eqnarray}
where $\kappa$ is the compressibility of the FL, $L(t) = \sqrt{D\tau}\propto \sqrt{\tau/ \rho}$ is the distance to which the excess charge can diffuse in time $\tau$, and $D$ is the diffusion constant of the FL.  In the Coulomb case,
\begin{equation}
E(\tau)=Q^2/L(\tau),
\end{equation}
where $L(\tau) = (e^2/h)\rho^{-1} t$.  Since, in both cases, $E(\tau)\to 1/\tau$ at long times, the resulting action is logarithmically divergent with system size,
\begin{eqnarray}
S_{hydro} && \sim \beta \log[L_{\infty}/L_B] \nonumber \\
\beta &&\propto  \rho[\Delta nL_B^2]^2.
\label{beta}
\end{eqnarray}
An estimate of $S_\epsilon\approx -\epsilon\tau_0$ is obtained by noting that the characteristic time involved in the charge motion is $\tau_0 = L_\infty^2/D$ in the dipolar case, and $\tau_0 = L_\infty(h/e^2)\rho$ in the Coulomb case.

The generalization of this analysis that leads to Eq. \ref{tij} is now obvious.  Replacing $L_\infty$ in $S_{hydro}$ with $R_{ij}\equiv |{\bf  R}_i - {\bf  R}_j|$ leads to the stated result.  Clearly,
\be
B \propto \exp[-2S_{melt}]
\label{B}
\ee
is exponentially small but finite, as promised.

There is an additional, potentially stronger (exponential) $R_{ij}$ dependence of $t_{ij}$ inherited from $S_{\epsilon}$ with $\epsilon\equiv \epsilon_i-\epsilon_j$, which we have ignored in this discussion.  However, since for the relevant pairs of sites, $|\epsilon_i-\epsilon_j| \sim \delta$ is small, and $R_{ij}$ is only large in proportion to $\delta^{-1/2}$, the product $|\epsilon_i-\epsilon_j| \tau_0\sim \delta^0$ in the dipolar case and $|\epsilon_i-\epsilon_j| \tau_0\sim \delta^{1/2}$ in the Coulomb case; in neither case does $S_{\epsilon}$ make an important contribution to the action.

Our calculations are controlled only
in the limit when $\Delta n L_{B}^{2}\gg 1$, the amplitude in Eq. \ref{B} is exponentially small, and consequently  the mobility of the bubbles at $T=0$ is correspondingly small.


\begin{thebibliography}{99}
\bibitem{note1} We would like to contrast these pahses
 with a system of domains
in 3D ferromagnets, where the energy stored in domain
 walls is not proportional to the sample volume and,
  therefore the domains do not constitute new phases.
\bibitem{He_3He_4} In support of this assumption we would like to mention the following experimental fact. Heat capacities of  two  other strongly correlated quantum liquids $^{3}$He and $^{4}$He at $T_{d}\ll T\ll \Omega_{D}$ have similar $T$-dependecies and comparable values. Here $T_{d}$ is a quantum degeneracy temperature (Fermi energy in $^{3}$He and condensation energy  in $^{4}$He.); and $\Omega_{D}$ is the Debye frequency, which plays in these
liquids the same role as the plasma frequency in electron liquid.
\bibitem{stripeQH}  Such a pattern has been predicted in quantum Hall devices \cite{fogler}, and observed in experiments Ref. \cite{eisentsteinqh}, where a giant and strongly $T$ and $H_\perp$ dependent resistance anisotropy has been observed in ultra-clean n-GaAs devices.   On a more general level this confirms the existence of a nematic symmetry breaking \cite{fradkinqh} in electronic systems.
\end{thebibliography}
\end{document}